\documentclass[twocolumn,showpacs,showkeys,amsmath,amssymb,nofootinbib,pra]
{revtex4-1}
     
\usepackage{color}
\usepackage{bm}
\usepackage{graphicx}  

\newtheorem{theorem}{Assertion} 	
 
\newcommand{\rd}{{\mathrm d}}
\newcommand{\re}{{\mathrm e}}
\newcommand{\ri}{{\mathrm i}}
\newcommand{\kL}{k_{\mathrm L}}
\newcommand{\ER}{E_{\mathrm R}}
\newcommand{\db}{\rangle\!\rangle}            
\begin{document}

\title[Quasienergy bands]
	{Floquet engineering with quasienergy bands 
	of periodically driven optical lattices}
	
\author{Martin Holthaus}
\email[e-mail: ]{martin.holthaus@uni-oldenburg.de}

\affiliation{Institut f\"ur Physik, 
	Carl von Ossietzky Universit\"at, 
	D-26111 Oldenburg, Germany}

\date{August 26, 2015}
 
\begin{abstract}
A primer on the Floquet theory of periodically time-dependent quantum systems
is provided, and it is shown how to apply this framework for computing the
quasienergy band structure governing the dynamics of ultracold atoms in 
driven optical cosine lattices. Such systems are viewed here as spatially and 
temporally periodic structures living in an extended Hilbert space, giving rise
to spatio-temporal Bloch waves whose dispersion relations can be manipulated 
at will by exploiting ac-Stark shifts and multiphoton resonances. The elements 
required for numerical calculations are introduced in a tutorial manner, and 
some example calculations are discussed in detail, thereby illustrating future 
prospects of Floquet engineering.     
\end{abstract} 

\pacs{67.85.Hj, 42.50.Hz, 32.80.Xx}


\keywords{Optical lattices, 
	        ultracold atoms, 
		quantum Floquet theory, 
		multiphoton transitions,
		spatio-temporal Bloch waves,
		Floquet engineering} 

\maketitle 


\section{What this is about}
\label{S_1}

The study of ultracold atoms in optical lattices by now has become a major 
branch of atomic physics~\cite{MorschOberthaler06,BlochEtAl08,LewensteinEtAl12}.
In particular, increasing effort is currently being devoted to exert a
controlling influence on atoms in optical lattices by subjecting them to a
time-periodic external force~\cite{SteckEtAl01,HensingerEtAl01,GemelkeEtAl05,
	LignierEtAl07,IvanovEtAl08,SiasEtAl08,EckardtEtAl09,ZenesiniEtAl09,
	AlbertiEtAl09,HallerEtAl10,StruckEtAl11,MaEtAl11,ChenEtAl11}.
Recent experiments in this fast-growing area have addressed the realization
and application of artificial tunable gauge potentials~\cite{StruckEtAl12,
	HaukeEtAl12,StruckEtAl13}, the realization of the Hofstadter 
Hamiltonian~\cite{AidelsburgerEtAl13,MiyakeEtAl13}, the observation of 
effective ferromagnetic domains~\cite{ParkerEtAl13}, the realization of 
the topological Haldane model~\cite{JotzuEtAl14}, and the creation of a 
roton-maxon dispersion for a Bose-Einstein condensate in a shaken optical
lattice~\cite{HaEtAl15}. These activities indicate that ultracold atoms in 
periodically driven optical lattices have a high potential for simulating a 
wide variety of condensed-matter systems and even models of high-energy 
physics, thereby offering new approaches to long-standing open questions.

The theoretical tool heavily used in this emerging new field is the Floquet
formalism. While this is well familiar to researchers working with atoms
and molecules in strong laser fields, it does not belong to the traditional
training of a cold-atoms physicist. This has led to a palpable knowledge gap: 
The Ph.D.~student or postdoctoral researcher performing experiments with 
driven optical lattices requires an easily accessible, sharply focused
introduction which should acquaint her or him with the prospects and pitfalls 
of the general concepts, giving advice how to perform one's own numerical 
simulations, and thus helping to obtain fresh ideas for specifically targeted 
further measurements.  

The present tutorial article is intended to meet just this demand. In contrast
to excellent recent review articles~\cite{GoldmanDalibard14,Eckardt15} which 
give a fairly general overview, and explain certain approximation schemes, it 
is essentially a manual on how to compute, and interprete, the quasienergy band
structure of a driven one-dimensional cosine lattice. Once the novice has 
mastered this, it will be found an easy task to adapt these methods to other 
situations of interest, such as more complicated lattice geometries, or 
different forms of driving. Besides, driven optical cosine lattices are likely 
to serve as {\em the\/} workhorses for Floquet engineering for years to come, 
so that the exclusive concentration on this one particular system appears to be
well justified. 

The material is organized as follows: In Sec.~\ref{S_2} we briefly review
the calculation of the energy bands of a stationary cosine lattice, in a form
that will be taken up again in Sec.~\ref{S_4}. Before, Sec.~\ref{S_3} offers 
a primer on the Floquet theory of periodically time-dependent quantum systems; 
this could be read independently of the other parts of this paper. 
There we consider a deceptively simple model, the driven particle in a box.
This allows us to introduce and discuss on an elementary level several facets 
of the Floquet picture which also govern, to a higher degree of sophistication, 
the physics of cold atoms in driven optical lattices. In Sec.~\ref{S_4} these 
tools are combined, and applied to  the determination of the spatio-temporal 
Bloch waves which form the backbones of the dynamics in time-periodically 
driven optical lattices. We explain the central elements required for numerical
computations, and discuss the results of some selected example calculations, 
hoping that these may inspire the reader to explore still further parameter 
regimes. A short outlook given in Sec.~\ref{S_5} concludes this tutorial. 

A word on referencing: There exists such a wealth of papers on diverse aspects 
of the Floquet formalism that it is impossible to do justice to all of them. 
The selection of references made here naturally carries a strong personal bias,
and I have to apologize to all colleagues who do not find their works properly 
cited. But given the availability of modern databases, a mere compilation of 
who did what might perhaps be found less useful than a certain pre-selection 
which provides a firm, definite view on the Floquet picture.

\section{Preliminaries: Band structure of a $1d$~cosine lattice}
\label{S_2}

\begin{figure}
\includegraphics[width = 0.9\linewidth]{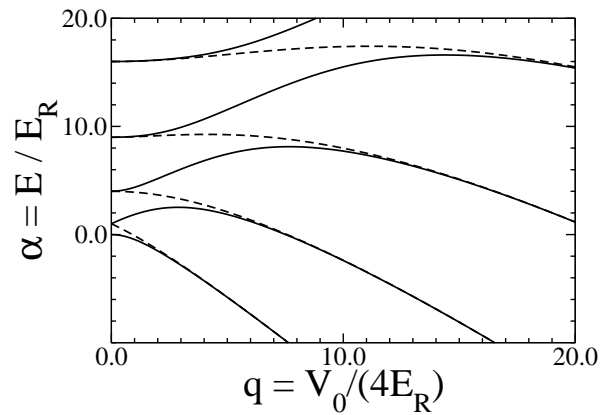}
\caption{Characteristic values $a_0$, $a_1$, $a_2$, $a_3$, $a_4$ (full lines,
	bottom to top) and $b_1$, $b_2$, $b_3$, $b_4$ (dashed lines, bottom 
	to top). For given scaled lattice depth $q = V_0/(4\ER)$, the $n$th 
	energy band of a $1d$ optical cosine lattice ranges from $a_n(q)$ to 
	$b_{n+1}(q)$, in units of the recoil energy~$\ER$.}
\label{F_1}	
\end{figure}

Let us consider a single quantum particle of mass~$M$ moving on a 
one-dimensional cosine lattice oriented along the $x$-axis, as described
by the Hamiltonian 
\begin{equation}
	H(x) = -\frac{\hbar^2}{2M} \frac{\rd^2}{\rd x^2} 
	+ \frac{V_0}{2}\cos(2\kL x) \; .
\end{equation}
Here $V_0$ denotes the lattice depth; in the particular case of an optical 
lattice $\kL$ is the wave number of the lattice-generating laser radiation. 
Our first goal is to solve the stationary Schr\"odinger equation
\begin{equation}
	H(x)\varphi(x) = E\varphi(x) \; ,	
\label{eq:EVE}
\end{equation}
and thereby to determine the band structure of this lattice; this problem
has been treated in great detail by Slater in the earlier days of solid-state
physics~\cite{Slater52}. To this end we introduce the dimensionless coordinate 
$z = \kL x$, so that the eigenvalue equation~(\ref{eq:EVE}) takes the form
\begin{equation}
	\left( -\frac{\hbar^2\kL^2}{2M} \frac{\rd^2}{\rd z^2}
	+ \frac{V_0}{2} \cos(2z) - E \right)\varphi(z) = 0 \; .
\end{equation}
Here we have sloppily but conveniently written $\varphi(z)$ instead of the 
mathematically correct $\varphi(z/\kL)/\sqrt{\kL}$. This step allows us to 
identify the quantity
\begin{equation}
	\ER = \frac{\hbar^2 \kL^2}{2M} \; 
\label{eq:REC}
\end{equation}
as the relevant energy scale of the problem; in case of an optical lattice 
this equals the familiar single-photon recoil energy of the 
particle~\cite{MorschOberthaler06}. Dividing, we 
obtain 
\begin{equation}
	\left( \frac{\rd^2}{\rd z^2} + \frac{E}{\ER}
	- 2 \frac{V_0}{4\ER}\cos(2z) \right) \varphi(z) = 0 \; .
\end{equation}	
This is precisely the standard form of the Mathieu 
equation~\cite{AbramowitzStegun70},
\begin{equation}
	\varphi''(z) + \big[ \alpha - 2q \cos(2z) \big] \varphi(z) = 0 \; ,
\label{eq:MAT}
\end{equation}
with parameters
\begin{eqnarray}
	\alpha & = & \frac{E}{\ER} \; ,
\\	
	q      & = & \frac{V_0}{4\ER} \; .
\end{eqnarray}		
Therefore, the famous stability chart of the Mathieu equation, which also 
determines the parameters of stable ion motion in a Paul trap~\cite{Paul90},
gives essential information on the states of a particle in an optical lattice. 
Namely, according to the Bloch theorem~\cite{Bloch29,AshcroftMermin76} the
solutions to the eigenvalue equation~(\ref{eq:EVE}) are Bloch waves
characterized by a wave number~$k$, 
\begin{equation}
	\varphi_k(x) = \re^{\ri k x} u_k(x) \; .
\label{eq:BLA}	
\end{equation}
In Bloch's own words, these are ``de Broglie waves which are modulated in the
rhythm of the lattice structure''~\cite{Bloch29}, since the functions $u_k(x)$
inherit the periodicity of the lattice potential: 
\begin{equation}
	u_k(x) = u_k(x + \pi/\kL) \; .
\label{eq:PBC}
\end{equation}
In terms of the scaled coordinate~$z$ this implies that we are looking for 
$\pi$-periodic solutions $\varphi(z) = \varphi(z + \pi)$ to the Mathieu
equation~(\ref{eq:MAT}) when considering a band edge $k/\kL = 0$. The 
opposite edges, where $k/\kL = \pm 1$, then give rise to solutions which 
change sign after one period, $\varphi(z) = -\varphi(z + \pi)$, and hence are 
$2\pi$-periodic. Now it is well known that for a given value of the Mathieu 
parameter~$q$, that is, for a given lattice depth, these desired periodic 
solutions exist only if the other parameter $\alpha$ adopts one of the 
discrete so-called characteristic values $a_r(q)$, which produce even Mathieu 
functions, or $b_r(q)$, which belong to odd functions; while even indices~$r$ 
designate $\pi$-periodic functions, odd indices refer to $2\pi$-periodic 
ones~\cite{AbramowitzStegun70}. Therefore, a plot of these characteristic 
values, such as depicted in Fig.~\ref{F_1}, immediately allows one to read off 
the width of the energy bands: For $n = 0,1,2\ldots$, the (energetically) lower
edge of the $n$th band is given by   
\begin{equation} 
	E_n^{\rm lower} = a_n(q) \ER \; , 
\end{equation}
while its upper edge is
\begin{equation}
 	E_n^{\rm upper} = b_{n+1}(q) \ER \; ;
\end{equation}	 	
note that the state with $k/\kL = 0$ alternates between the lower and the 
upper edge from band to band.

In order to compute the full dispersion relations $E_n(k)$ of the cosine
lattice we insert the Bloch ansatz~(\ref{eq:BLA}) into the eigenvalue
equation~(\ref{eq:EVE}), employing the momentum operator  
\begin{equation}
	p = \frac{\hbar}{\ri} \frac{\rd}{\rd x} \; ,
\end{equation}	
thus obtaining
\begin{equation}
	\left( \frac{p^2}{2M} + \frac{V_0}{2}\cos(2\kL x) \right)
	\re^{\ri k x} u_k(x) = E(k) \re^{\ri k x} u_k(x) \; .
\label{eq:NPR}	
\end{equation}
This form of the eigenvalue problem is not practical, because here the
eigenfunctions consist of both the plane-wave factors $\re^{\ri k x}$ and 
the Bloch functions $u_k(x)$, while one has to impose the periodic boundary 
condition~(\ref{eq:PBC}) on the latter only. Therefore, in the spirit of the 
$k \cdot p$-approach to band structures~\cite{Harrison89} one multiplies this 
Eq.~(\ref{eq:NPR}) from the left by $\re^{-\ri k x}$ and utilizes the identity 
\begin{eqnarray}
	\re^{-\ri k x} p \re^{\ri k x} & = & p - \ri k \big[x,p\big]
\nonumber \\	& = &
	p + \hbar k \; ,
\label{eq:BCH}
\end{eqnarray}
arriving at
\begin{equation}
	\left( \frac{(p + \hbar k)^2}{2M} + \frac{V_0}{2}\cos(2\kL x) \right)
	u_k(x) = E(k) u_k(x) \; .
\end{equation}	
Multiplying out the squared momenta, this gives 
\begin{eqnarray}
	& & 
	\left( \frac{p^2}{2M} + \frac{\hbar k \cdot p}{M}
	+ \frac{\hbar^2 k^2}{2M} + \frac{V_0}{2}\cos(2\kL x) \right) u_k(x)
\nonumber \\	 
	& = & E(k) u_k(x) \; .
	\phantom{\frac{p^2}{2M}}
\end{eqnarray}	
This modified eigenvalue equation for the Bloch functions alone is much easier 
to deal with than its antecessor~(\ref{eq:NPR}), because one now can expand the
eigenfunctions with respect to a suitable basis which already incorporates the 
periodic boundary conditions, then represent all operators in this basis, and 
diagonalize the resulting matrices numerically. Specifically, upon return to 
the dimensionless coordinate $z = \kL x$ and division by $\ER$ we have 
\begin{eqnarray}
	& &
	\left( - \frac{\rd^2}{\rd z^2} + 2 \frac{k}{\kL} 
	\frac{1}{\ri} \frac{\rd}{\rd z} + \left(\frac{k}{\kL}\right)^2
	+ \frac{V_0}{2\ER}\cos(2z) \right) u_k(z)
\nonumber \\	 
	& = & \frac{E(k)}{\ER} u_k(z) \; ,
\label{eq:EVP}
\end{eqnarray}	
requiring $\pi$-periodic solutions
\begin{equation}
	u_k(z) = u_k(z + \pi) \; .
\end{equation}	
In order to satisfy these boundary conditions we simply choose the basis  
$\{ \varphi_\mu(z) \; ; \; \mu = 0,1,2,3,\ldots \}$ of normalized 
$\pi$-periodic trigonometric functions, so that
\begin{equation}
	\varphi_0(z) = \sqrt{\frac{1}{\pi}}
\label{eq:BA0}
\end{equation}
and
\begin{equation}
	\varphi_\mu(z) = \left\{
	\begin{array}{ll}
	\displaystyle
	\sqrt{\frac{2}{\pi}}\sin\big([\mu+1]z\big) 
	& ; \; \mu = 1,3,5,\ldots	\\
	\displaystyle
	\sqrt{\frac{2}{\pi}}\cos( \mu z) 
	& ; \; \mu = 2,4,6,\ldots \; .\\
	\end{array}
	\right.
\label{eq:BAR}		
\end{equation}	
In this basis the negative second derivative is respresented by a diagonal 
matrix with quadratically-growing entries,
\begin{equation}
	-\frac{\rd^2}{\rd z^2} = 
	\left(
	\begin{array}{cccccc}
	0 & & &  &  &	\\
	& 4 & &  &  &	\\
	& & 4 &  &  & 	\\
	& & & 16 &  &	\\
	& & & &  16 & 	\\
	& & & &  &  \ldots \\
	\end{array}
	\right) \; , 
\label{eq:MX1}
\end{equation}
whereas the first derivative leads to a matrix with entries on the first
off-diagonal only,
\begin{equation}
	\frac{\rd}{\rd z} =
	\left(
	\begin{array}{ccrcrc}
	0 &   &    &   &    &		\\
	  & 0 & -2 &   &    &    	\\
	  & 2 &  0 &   &    &     	\\
	  &   &    & 0 & -4 &    	\\
	  &   &    & 4 &  0 & 	 	\\
	  &   &    &   &    & \ldots 	\\
	\end{array} 
	\right) \; .
\label{eq:MX2}	 
\end{equation}
Finally, the cosine potential has non-vanishing matrix elements only on
the second off-diagonal,
\begin{equation}
	\cos(2z) = \frac{1}{2} 
	\left(
	\begin{array}{ccccccc}
	0        & 0 & \sqrt{2} &   &   &   &	\\
	0        & 0 &       0  & 1 &   &   & 	\\
	\sqrt{2} & 0 &       0  & 0 & 1 &   &   \\
	         & 1 &       0  & 0 & 0 & 1 &	\\
	         &   &       1  & 0 & 0 & 0 & \ldots	\\
	         &   &          & 1 & 0 & 0 &	\\
		 &   &          &   & \ldots &   & 
	\end{array}
	\right) \; .
\label{eq:MX3}	 
\end{equation}
In all these matrices~(\ref{eq:MX1}), (\ref{eq:MX2}), and~(\ref{eq:MX3}), 
elements not shown are zero. Given the lattice depth $V_0/\ER$, one can 
then for any $k/\kL$ set up the matrix corresponding to the operator on the 
left-hand side of Eq.~(\ref{eq:EVP}), and diagonalize, making sure that the 
matrix size is chosen sufficiently large so that the desired eigenvalues are 
converged to the accuracy specified.

\begin{figure}
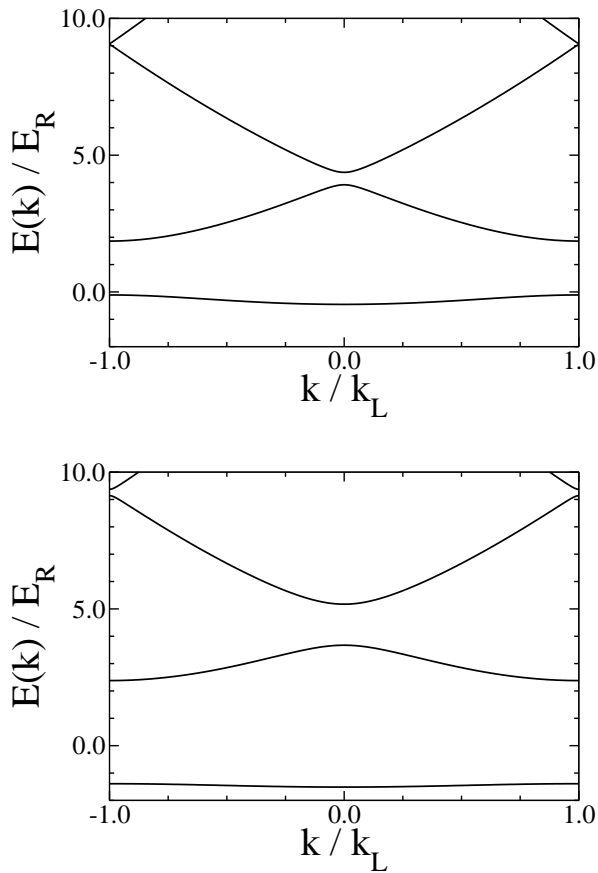

\includegraphics[width = 0.9\linewidth]{QES_Fig2a.eps}

\vspace*{4ex}

\includegraphics[width = 0.9\linewidth]{QES_Fig2b.eps}
\caption{Dispersion relations $E_n(k)$ for the lowest three bands $n = 0,1,2$
	of a cosine lattice with depth $V_0/\ER = 4.0$ (upper panel) or 
	$V_0/\ER = 8.0$ (lower panel), restricted to the first Brillouin zone. 
	Observe the widening of the gaps with increasing lattice depth.}
\label{F_2}	
\end{figure}

In Fig.~\ref{F_2} we show the dispersion relations $E_n(k)$ for the lowest 
three bands $n = 0,1,2$ of an optical cosine lattice with depth 
$V_0/\ER = 4.0$ and $V_0/\ER = 8.0$, respectively; these two examples will 
be taken up again in Sec.~\ref{S_4} for illustrating the engineering options 
offered by a time-periodic force. Since the lattice period  is $a = \pi/\kL$, 
the first Brillouin zone $-\pi/a < k \leq +\pi/a$ here ranges from 
$-\kL$ to $+\kL$ in $k$-space; if necessary, the dispersion relations 
outside this interval are obtained by periodic continuation, 
$E_n(k) = E_n(k + 2\kL)$~\cite{AshcroftMermin76}.

\section{Tools: Basic elements of the Floquet picture}
\label{S_3}

We now put together some elements which are required for the description of
quantum systems possessing a discrete translational invariance in time 
rather than in space. In essence, these go back to the work of the French
mathematician Achille Marie Gaston Floquet (1847 - 1920) on linear differential
equations with periodic coefficients~\cite{Floquet83}, as does Bloch's analysis
of the quantum mechanics of electrons in crystal lattices~\cite{Bloch29}. But
there are some peculiarities of time-periodic quantum systems which, although
they are known in principle~\cite{AutlerTownes55,Shirley65,Zeldovich66,
Ritus66,Sambe73,FainshteinEtAl78,GrifoniHaenggi98,ChuTelnov04,Howland92}, 
appear to be rarely appreciated, so that this condensed summary may be found 
useful. Since there are no principal differences between single-particle and 
many-body systems on this level, we keep this account fairly general.

\subsection{Mathematical foundation}
\label{SubSec31}

Thus, we consider a quantum system defined on a Hilbert space ${\mathcal H}$
and governed by a Hamiltonian which is periodic in time with period~$T$,  
\begin{equation}
	H(t) = H(t + T) \; ,
\label{eq:PTH}	
\end{equation}
and we aim at exploring the consequences of this periodicity for the solutions 
to the time-dependent Schr\"odinger equation, 
\begin{equation}
	\ri\hbar\frac{\rd}{\rd t} | \psi(t) \rangle 
	= H(t) | \psi(t) \rangle \; .
\label{eq:SGL}
\end{equation}
Here we employ the abstract bra-ket notation for quantum states, so that, 
for instance, $\psi(x,t) = \langle x | \psi(t) \rangle$. Instead of 
solving Eq.~(\ref{eq:SGL}) for one particular inital condition we desire a 
characterization of {\em all\/} its solutions, so that we have to construct 
the unitary time-evolution operator $U(t,0)$ which effectuates the propagation 
of any initial state $|\psi(0)\rangle$ in time~\cite{Sakurai94},
\begin{equation}
	| \psi(t) \rangle = U(t,0) | \psi(0) \rangle \; .
\end{equation}
This operator $U(t,0)$ itself obeys the Schr\"odinger-like equation   
\begin{equation}
	\ri\hbar \frac{\rd}{\rd t} U(t,0) = H(t) U(t,0)
\label{eq:SLE}
\end{equation}		
with the initial condition
\begin{equation}
	U(0,0) = {\rm id} \; ,
\end{equation}	
where $ {\rm id} $ denotes the identity operator on ${\mathcal H}$. For any 
Hamiltonian $H(t)$ the time-evolution operator has the physically transparent 
semigroup property
\begin{equation}
	U(t_1+t_2,0) = U(t_1+t_2,t_1) U(t_1,0) \; .
\end{equation}	
But in the particular case of a periodically time-de\-pen\-dent 
Hamiltonian~(\ref{eq:PTH}) one has even more:
\begin{theorem}
	If the Hamiltonian $H(t) = H(t+T)$ is periodic in time with period~$T$,
	the associated time-evolution operator $U(t,0)$ obeys the identity
\begin{equation}
	U(t+T,0) = U(t,0) U(T,0) \; .
\label{eq:BTM}
\end{equation}
\end{theorem}
This says that knowledge of $U(t,0)$ for $ 0 \leq t \leq T$ suffices to
construct $U(t,0)$ for all times~$t \ge 0$. Although this statement may appear 
``somewhat obvious'', it still needs to be proven. Fortunately, this is not 
difficult: Consider the composite operator
\begin{equation}
	V(t) := U(t+T,0) U^{-1}(T,0) \; .
\end{equation}	
Then, obviously, one has both
\begin{equation}
	V(0) = {\rm id} = U(0,0)
\end{equation}
and
\begin{eqnarray}
	\ri\hbar\frac{\rd}{\rd t} V(t) & = &
	\ri\hbar\frac{\rd}{\rd t} U(t+T,0) U^{-1}(T,0)
\nonumber \\	& = &	
	H(t+T) U(t+T,0) U^{-1}(T,0)
	\phantom{\frac{\rd}{\rd t}}
\nonumber \\	& = &   
	H(t) V(t) \; . 
	\phantom{\frac{\rd}{\rd t}}
\end{eqnarray}
Thus, $U(t,0)$ and $V(t)$ obey the same differential equation with the same
initial condition, and therefore coincide, which proves the assertion. 
\hfill $\Box$ 

In the following we restrict ourselves temporarily to systems with a 
finite-dimensional Hilbert space ${\mathcal H}$; this restriction eliminates 
certain technical subtleties which will be discussed later with the help of 
a simple example. Just as the translation operator by a lattice vector plays 
a prominent role in solid-state physics, it is intuitively clear that the 
one-cycle evolution operator $U(T,0)$, which is also known as monodromy 
operator in the mathematical literature~\cite{DaleckiiKrein74}, must be 
of particular importance here. We write this one-cyle evolution operator as 
an exponential in the suggestive form   	
\begin{equation}
	U(T,0) = \exp(-\ri G T/\hbar) \; ,
\end{equation}	
where the operator~$G$ is Hermitian, possessing real eigenvalues: This makes 
sure that $\exp(-\ri G T/\hbar)$ is unitary, so that all its eigenvalues lie
on the unit circle. With the help of this exponential we now define a further 
unitary operator: 
\begin{equation}
	P(t) := U(t,0) \exp(+\ri G t/\hbar) \; .
\label{eq:DFP}
\end{equation}
Then one deduces
\begin{eqnarray}
	P(t+T) & = & U(t+T,0) \exp\big(+\ri G (t + T)/\hbar\big)
\nonumber \\	& = &
	U(t,0) \Big( U(T,0) \exp(+\ri G T / \hbar) \Big) \exp(+\ri G t / \hbar)
\nonumber \\	& = &	
	P(t) \; ,
\label{eq:PPP}
\end{eqnarray}	
where the basic assertion~(\ref{eq:BTM}) has been used in the first step, and 
the definition~(\ref{eq:DFP}) in the second. Thus, we can formulate a second
important insight:
\begin{theorem}
	Under the propositions specified above, the time-evolution operator
	$U(t,0)$ of a $T$-periodically time-dependent quantum system has the 
	form
\begin{equation}
	U(t,0) = P(t) \exp(-\ri G t/\hbar) \; ,
\label{eq:FRU}
\end{equation}
	where the unitary operator $P(t) = P(t + T)$ is $T$-periodic, and
	the operator~$G$ is Hermitian.
	\hfill $\Box$	 		   
\end{theorem}	
Writing the set of eigenvalues of $U(T,0) = \exp(-\ri G T/\hbar)$ as
$\{\re^{-\ri\varepsilon_n T/\hbar}\}$, and its eigenstates as 
$\{| n \rangle \}$, we have a spectral representation
\begin{equation}
	U(T,0) = \sum_n 
	| n \rangle \re^{-\ri\varepsilon_n T/\hbar} \langle n | \; ,
\label{eq:SRU}
\end{equation}	
implying
\begin{equation}
	\re^{-\ri G t/\hbar} | n \rangle = 
	\re^{-\ri \varepsilon_n t/\hbar} | n \rangle \; .
\end{equation}
Now we are in a position to monitor the time-evolution of an arbitrary
initial state $| \psi(0) \rangle$: Expanding with respect to the eigenstates 
of $U(T,0)$, we start with
\begin{eqnarray}
	| \psi(0) \rangle & = & \sum_n | n \rangle \langle n | \psi(0) \rangle 	 
\nonumber \\ 	& = &
	\sum_n a_n | n \rangle ,
\end{eqnarray}
where $a_n = \langle n | \psi(0) \rangle $. Applying $U(t,0)$, we then find 
\begin{eqnarray}
	| \psi(t) \rangle & = & U(t,0) | \psi(0) \rangle
	\phantom{\sum_n}
\nonumber \\	& = &
	\sum_n a_n P(t) \re^{-i G t/\hbar} | n \rangle
\nonumber \\	& = &
	\sum_n a_n P(t) | n \rangle \re^{-\ri \varepsilon_n t/\hbar}
\nonumber \\	& = &
	\sum_n a_n | u_n(t) \rangle \re^{-\ri \varepsilon_n t/\hbar} \; .		
\label{eq:TEV}
\end{eqnarray}
In the last step made here we have {\em defined\/} the Floquet functions
\begin{equation}
	| u _n(t) \rangle := P(t) | n \rangle \; ,
\label{eq:FLF}
\end{equation}	 
which, as a consequence of the identity~(\ref{eq:PPP}), are $T$-periodic:
\begin{equation}
	| u_n(t) \rangle = | u_n(t + T) \rangle \; .
\end{equation}
For the sake of definite nomenclature we will refer to the states
\begin{equation}
	| \psi_n(t) \rangle = | u_n(t) \rangle 
	\re^{-\ri\varepsilon_n t/\hbar}
\label{eq:FST}
\end{equation}
as {\em Floquet states\/}; note that these states, in contrast to the 
$T$-periodic Floquet functions $|u_n(t)\rangle$, are solutions to the 
time-dependent Schr\"odinger equation~(\ref{eq:SGL}). Since the normalized
eigenfunctions $\{ | n \rangle \}$ of $U(T,0)$ form a complete set, and $P(t)$ 
is unitary, so do the Floquet functions~(\ref{eq:FLF}) at each instant~$t$. 
Thus, we can formulate the content of Eq.~(\ref{eq:TEV}) as follows:
\begin{theorem}
	Under the propositions specified above, any solution 
	$| \psi(t) \rangle$ to the time-dependent Schr\"odinger 
	equation~(\ref{eq:SGL}) with a $T$-periodic Hamiltonian 
	$H(t)$ can be expanded with respect to the Floquet states,
\begin{equation}
	| \psi(t) \rangle = \sum_n a_n | u_n(t) \rangle 
	\re^{-\ri \varepsilon_n t/\hbar} \; ,		
\label{eq:FEX}
\end{equation}	
	where the coefficients $a_n$ do {\em not} depend on time. 
	\hfill $\Box$
\end{theorem}
The last half-sentence is of central importance for countless applications of 
this Floquet picture: Since the periodic time-dependence is already 
incorporated into the basis, the expansion coefficients remain constant. This 
implies that one can assign occupation probabilities $| a_n |^2$ to the Floquet states 
which are preserved despite the action of the time-periodic influence, so that 
several concepts and techniques used for time-independent quantum systems can 
be carried over to periodically time-dependent ones. Indeed, the phase factors 
$\re^{-\ri\varepsilon_n t/\hbar}$ showing up in this expansion~(\ref{eq:FEX}) 
resemble the factors $\re^{-\ri E_n t/\hbar}$ which accompany the 
time-evolution of energy eigenstates with energies~$E_n$ if their Hamiltonian 
does not depend on time: The quantities $\varepsilon_n$ look as if they were 
energies, and therefore are aptly named {\em quasienergies\/}; this designation
appears to have been coined in 1966 almost simultaneously by the eminent 
Soviet physicists Yakov Borisovich Zel'dovich~\cite{Zeldovich66} and Vladimir 
Ivanovich Ritus~\cite{Ritus66}.

\subsection{The Brillouin-zone structure of the quasienergy spectrum}
\label{SubSec32}

In a formal sense, it seems tempting to interprete the preceding considerations
as follows: Perform the unitary transformation
\begin{equation}
	| \psi(t) \rangle = P(t) | \widetilde{\psi}(t) \rangle \; ,
\label{eq:UTR}
\end{equation}	 
so that
\begin{equation}
	\ri \hbar\frac{\rd}{\rd t} | \psi(t) \rangle 
	= \ri \hbar \dot{P}(t) | \widetilde{\psi}(t) \rangle
	+ P(t) \ri \hbar\frac{\rd}{\rd t} | \widetilde{\psi}(t) \rangle \; ,
\end{equation}	
where the overdot means differentiation with respect to~$t$. Now the 
definition~(\ref{eq:DFP}) implies  
\begin{equation}
	\ri \hbar \dot{P}(t) = \ri \hbar \dot{U}(t) \exp(\ri G t/\hbar)
	- U(t) G \exp(\ri G t/\hbar) \; , 
\end{equation}	
giving
\begin{eqnarray}
	\ri \hbar\frac{\rd}{\rd t} | \psi(t) \rangle 
	& = & H(t) U(t) \exp(\ri G t/\hbar) | \widetilde{\psi}(t) \rangle
\nonumber \\ 
	& - & U(t) G \exp(\ri G t/\hbar) | \widetilde{\psi}(t) \rangle
\nonumber \\ 
	& + & P(t) \ri \hbar\frac{\rd}{\rd t} | \widetilde{\psi}(t) \rangle
	\; ,
\label{eq:UFG}
\end{eqnarray}
where Eq.~(\ref{eq:SLE}) has been used. Next, the representation~(\ref{eq:FRU})
together with the defining Eq.~(\ref{eq:UTR}) readily yields	
\begin{equation}
	U(t) \exp(\ri G t/\hbar) | \widetilde{\psi}(t) \rangle
	= | \psi(t) \rangle \; , 
\end{equation}	
so that the left-hand side of Eq.~(\ref{eq:UFG}) cancels the first term on 
the right-hand side, leaving us with
\begin{eqnarray}
	\ri \hbar\frac{\rd}{\rd t} | \widetilde{\psi}(t) \rangle 
	& = &
	P^{-1}(t) U(t) \exp(\ri G t/\hbar) G | \widetilde{\psi}(t) \rangle
\nonumber \\	& = &	
	G | \widetilde{\psi}(t) \rangle \; .
\label{eq:SEG}		
\end{eqnarray}	
This looks interesting: The time-independent operator $G$ here plays the role
of the Hamiltonian for the transformed states $| \widetilde{\psi}(t) \rangle$.
Thus, if the transformation~(\ref{eq:UTR}) corresponds to a change of the
frame of reference, it works such that the dynamics, as seen from the new 
reference frame, are governed by a time-independent Hamiltonian. In fact, 
this is the guiding principle behind the construction of the Floquet solutions 
to certain {\em integrable\/} periodically time-dependent problems, such as 
a two-level system in a circularly polarized classical radiation 
field~\cite{RabiEtAl54}: In that paradigmatically important example, which 
will also play a central role for understanding the particular optical-lattice 
engineering outlined in Chap.~\ref{SubSec45}, the Hamiltonian takes the form
\begin{equation}
	H_c(t) = \frac{\hbar\omega_0}{2} \sigma_z + \frac{\mu F}{2}
	\big( \sigma_x \cos \omega t + \sigma_y \sin \omega t \big) \; ,
\label{eq:TLC}
\end{equation}	
where $\sigma_{x,y,z}$ are the usual Pauli matrices~\cite{Sakurai94}, and 
$\omega_0$ denotes the frequency of a transition between the two states of 
the undriven system $H_0 = \hbar\omega_0\sigma_z/2$. The parameters $F$ and 
$\omega$ specify, respectively, the strength and the frequency of the driving 
force, and $\mu$ is the dipole matrix element connecting the two eigenstates
of~$H_0$. Then the transformation~(\ref{eq:UTR}) is designed such that it 
brings us to a frame of reference which co-rotates with the circularly 
polarized field:\footnote{The transformation to the co-rotating frame is 
	already achieved by the operator $\exp(-\ri\omega t \sigma_z/2)$. 
	The additional factor $\exp(\ri\omega t/2) \bm{1}$ is incorporated 
	here to ensure the required periodicity~(\ref{eq:PPP}).}         
\begin{equation}
	P(t) = \exp\big(\ri\omega t (\bm{1} - \sigma_z) /2 \big) \; ,
\label{eq:PTL}
\end{equation}
where $\bm{1}$ is the $2 \times 2$ unit matrix. Evidently, when viewed
from the co-rotating frame, the circularly polarized field must appear as a
time-independent one. Indeed, working out the transformation  
\begin{eqnarray}	& &
	P^{\dagger}(t)
	\left( H_c(t) - \ri\hbar\frac{\rd}{\rd t} \right)
	P(t)
\nonumber\\ & = &
	\frac{\hbar\omega}{2} \bm{1} +
	\frac{\hbar}{2}( \omega_0 - \omega ) \sigma_z
	+ \frac{\mu F}{2} \sigma_x - \ri\hbar\frac{\rd}{\rd t} 
\end{eqnarray}
allows one to identify the desired time-independent operator
\begin{equation}
	G_c = \frac{\hbar\omega}{2} \bm{1} +
	\frac{\hbar}{2}( \omega_0 - \omega ) \sigma_z
	+ \frac{\mu F}{2} \sigma_x \; ,
\label{eq:OGC}
\end{equation}			
which, upon diagonalization, yields the quasienergies
\begin{equation}
	\varepsilon_{\pm} = \frac{\hbar}{2}( \omega \pm \Omega )
\label{eq:TLQ}	
\end{equation}
with the generalized Rabi frequency
\begin{equation}
	\Omega = \sqrt{ (\omega_0 - \omega)^2 + (\mu F/\hbar)^2 } \; .	
\label{eq:GRF}
\end{equation}
Moreover, transforming the eigenstates of $G_c$ back to the laboratory frame
provides the Floquet states of the system~(\ref{eq:TLC}). These results are
also useful in an approximate sense when the driving field is linearly 
polarized, rather than circularly, so that the Hamiltonian reads
\begin{equation}
	H_l(t) = \frac{\hbar\omega_0}{2} \sigma_z + \mu F
	\sigma_x \cos \omega t \; .
\label{eq:TLL}
\end{equation}	
Now the linearly polarized field may be regarded as a superposition of two
circularly polarized components with opposite sense of rotation. If one 
transforms to a frame co-rotating with one of these components, that 
component appears stationary, whereas the other, counter-rotating component
acquires twice its original frequency. Formally this is expressed as  
\begin{eqnarray}	& &
	P^{\dagger}(t)
	\left( H_l(t) - \ri\hbar\frac{\rd}{\rd t} \right)
	P(t)
\nonumber\\ & = &
	G_c - \ri\hbar\frac{\rd}{\rd t} + \frac{\mu F}{2}
	\big( \sigma_x \cos 2\omega t - \sigma_y \sin 2\omega t \big) \; , 
\label{eq:TRA}
\end{eqnarray}
where the transformation $P(t)$ again is given by Eq.~(\ref{eq:PTL}). If 
one now neglects the double-frequency counter-rotating component, hoping 
that its effects will average out, quasienergies and Floquet states of the 
linearly forced two-level system~(\ref{eq:TLL}) again are provided by the 
operator~(\ref{eq:OGC}). This is the famous {\em rotaing-wave approximation\/} 
(RWA), which obviously constitutes a particular high-frequency 
approximation~\cite{AllenEberly88}.

To give another example of an integrable periodically time-dependent system 
which can be solved by means of a transformation~(\ref{eq:UTR}) we mention 
the linearly forced harmonic oscillator: Here the Floquet states are found  
by transforming to a reference frame attached to an oscillating periodic 
solution to the corresponding classical equation of 
motion~\cite{BreuerHolthaus89a,LangemeyerHolthaus14}. But in general, the
remarkable equation~(\ref{eq:SEG}) appears to good to be true. So where is 
the catch?  

A first piece of the answer already appears when taking the limit of the 
quasienergies~(\ref{eq:TLQ}) for vanishing driving amplitude, 
$\mu F/(\hbar\omega) \to 0$: Then these quasienergies do {\em not\/} reduce
to the energy eigenvalues $\pm \hbar\omega_0/2$ of the undriven two-level
system. Instead, one has to distinguish two cases: If $\omega < \omega_0$,
so that the driving frequency is detuned to the red side of the transition,
one finds
\begin{eqnarray}
	\varepsilon_+ & \to & +\hbar\omega_0/2 
\nonumber \\
	\varepsilon_- & \to & -\hbar\omega_0/2 + \hbar\omega \; .	 
\label{eq:LI1}
\end{eqnarray}
On the other hand, if the driving frequency is blue-detuned, meaning
$\omega > \omega_0$, one has
\begin{eqnarray}
	\varepsilon_+ & \to & -\hbar\omega_0/2 + \hbar\omega 
\nonumber \\
	\varepsilon_- & \to & +\hbar\omega_0/2 \; .	 
\label{eq:LI2}
\end{eqnarray}
In either case, what does the additional ``$+\hbar\omega$'' mean?   
     
As the alert reader will have noted when following the general reasoning
in the previous Chap.~\ref{SubSec31}, the {\em Floquet multipliers\/} 
$\{ \re^{-\ri\varepsilon_n T/\hbar} \}$ are well defined, being the eigenvalues
of the one-cycle evolution operator $U(T,0)$. However, these quantities are 
just complex numbers on the unit circle which have been parametrized in this 
particular manner to enforce the suggestive form~(\ref{eq:FST}) of the 
fundamental Floquet solutions, but they do not uniquely determine the 
quasienergies $\varepsilon_n$: The complex logarithm, which is needed to 
extract these quasienergies from the Floquet multipliers, is multi-valued. 
Since $\re^z = \re^{z + m2\pi\ri}$, where $m = 0,\pm 1,\pm 2, \ldots$ is an 
arbitrary integer, the Floquet multipliers thus fix the quasienergies only up 
to an integer multiple of $2\pi\hbar/T$. Introducing the angular frequency
\begin{equation}   
	\omega = \frac{2\pi}{T} \; ,
\label{eq:DFO}
\end{equation}
a quasienergy, labeled by the state index~$n$, should therefore be regarded 
as an entire {\em class\/}	
\begin{equation}
	\varepsilon_{(n,m)} := \varepsilon_n + m \hbar\omega
	\quad ; \quad m = 0, \pm 1, \pm 2, \ldots
\label{eq:QEC}
\end{equation}
{\em of equivalent representatives\/}, where 
$\varepsilon_n = \varepsilon_{(n,0)}$ has to be selected by some suitable 
convention; according to Eq.~(\ref{eq:QEC}), the quasienergy representative 
labeled $(n,m)$ then differs from that $\varepsilon_n$ by $m\hbar\omega$. 
For instance, $\varepsilon_n$ might be that representative which falls into 
the {\em first quasienergy Brillouin zone\/} 
$-\hbar\omega/2 < \varepsilon \leq +\hbar\omega/2$, but sometimes other 
choices are more useful, as in the example discussed in the following
Chap.~\ref{SubSec33}. This notion of a quasienergy Brillouin zone emphasizes 
the solid-state analogy: Just as a quasimomentum of a particle in a periodic 
potential $V(x) = V(x+a)$ is determined only up to $\hbar$ times an integer 
multiple of the reciprocal lattice ``vector'' $2\pi/a$, a quasienergy of a 
system governed by a $T$-periodic Hamiltonian $H(t) = H(t+T)$ is determined 
only up to an integer multiple of the ``photon'' energy $\hbar\omega$. 

Hence, the previous result~(\ref{eq:TLQ}) for the quasienergies of a driven
two-level system should be written more carefully as
\begin{equation}
	\varepsilon_{\pm} = \frac{\hbar}{2}( \omega \pm \Omega )
	\qquad \bmod \hbar\omega \; ,
\label{eq:QTM}
\end{equation}
thus resolving the question posed after inspecting the limits~(\ref{eq:LI1})
and~(\ref{eq:LI2}). Returning to the general case, a well-meant attempt to go 
beyond Eq.~(\ref{eq:SRU}) and to ``define'' an operator~$G$ acting on the same
Hilbert space ${\mathcal H}$ as the system's Hamiltonian $H(t)$ according to
\begin{equation}
	\mbox{`` }G = \sum_n |n \rangle \varepsilon_n \langle n |\mbox{ ''}
\label{eq:NAT}
\end{equation}
would be incomplete without an additional specification, either explicit 
or implicit, how to resolve the multi-valuedness of each individual
$\varepsilon_n$. What is more, in many cases it is not even desirable to 
single out one particular representative of a quasienergy class~(\ref{eq:QEC}),
because it is precisely the ``$\bmod \; \hbar\omega$-indeterminacy'' which 
allows for a physically most transparent description of multiphoton 
transitions induced by a periodic drive.

\subsection{Case study: ac-Stark shifts and multiphoton resonances}
\label{SubSec33}

In order to illustrate this important fact, and thereby to prepare the 
discussion of the physics occurring in driven optical lattices, we now 
undertake a small digression and consider the seemingly simple model of a 
``periodically driven particle in a box''~\cite{BreuerEtAl88}, which also 
prompts us to address some practical issues relevant for numerical computation:
A particle of mass~$M$ is supposed to move on the $x$-axis between hard walls 
located at $x = \pm a$, as modeled by the unpertubed Hamiltonian   	       
\begin{equation}
	H_0(x) = \frac{-\hbar^2}{2M} \frac{\rd^2}{\rd x^2} + V(x)
\label{eq:UPO}
\end{equation}
with the archetypal ``box'' potential
\begin{equation}
	V(x) = \left\{	\begin{array}{rl}
		0 	& , \quad | x | < a	\\
		\infty 	& , \quad | x | \ge a 	\\
			\end{array} \right.
\end{equation}			
which forces the particle's wave function to vanish at $x = \pm a$. Moreover, 
the particle is subjected to a sinusoidal force with angular frequency
$\omega$ and amplitude $F_0$, conforming to the total Hamiltonian
\begin{equation}
	H(x,t) = H_0(x) - F_0 x \cos(\omega t) \; .
\label{eq:PIB}
\end{equation}	
For demonstration purposes we fix the driving frequency such that it is 
red-detuned by $5\%$ from the dipole-allowed transition between the unperturbed
box ground state with energy~$E_1$, and the first excited state with energy 
$E_2 = 4 E_1$:    
\begin{equation}
	\hbar\omega = 0.95 \, (E_2 - E_1) \; .
\label{eq:RLF}
\end{equation}
Since the model lives in an infinite-dimensional Hilbert space ${\mathcal H}$, 
it is actually {\em not\/} fully covered by the mathematical assertions 
formulated in Chap.~\ref{SubSec31}. But since the energies~$E_n$ of the 
unperturbed box states~$\varphi_n(x)$ grow quadratically with their quantum 
number~$n$, 
\begin{equation}
	E_n = \frac{\hbar^2 \pi^2}{8Ma^2} \; n^2 
	\quad ; \quad n = 1, 2, 3, \ldots \; ,
\end{equation}
and the dipole matrix element conecting the states $\varphi_n(x)$ and
$\varphi_m(x)$ falls off quite fast when the difference $|m - n|$ becomes
large, 
\begin{equation}
	\langle \varphi_m | x | \varphi_n \rangle =
	\left\{ \begin{array}{ll}
	- \displaystyle\frac{16a}{\pi^2}\frac{mn}{(m^2 - n^2)^2}  
		&, \; m + n \text{ odd}  \\
	\phantom{\displaystyle\sum} 0 	\
		&, \; m + n \text{ even} \; ,
	\end{array} \right.
\label{eq:DME}
\end{equation}	
it seems reasonable to assume that an external force with the relatively low 
frequency~(\ref{eq:RLF}), and with moderate strength, will hardly affect the 
high-lying states. We therefore truncate the Hilbert space, retaining only the 
subspace spanned by the $n_{\rm max}$ lowest energy eigenstates~$\varphi_n(x)$ 
of the unperturbed operator~(\ref{eq:UPO}), and then have to deal with a 
system of $n_{\rm max}$ complex coupled ordinary differential equations which 
can be integrated numerically by standard routines. In order to calculate the 
truncated time-evolution matrix $U(T,0)$ one takes each of the unperturbed box 
eigenstates as initial condition, $\psi_n(x,0) = \varphi_n(x)$, and computes 
the states $\psi_n(x,T)$ resulting after one period, collecting these state 
vectors as columns of the mono\-dromy matrix. Finally, diagonalizing this 
matrix yields the finite-subspace approximations to the Floquet multipliers 
$\{ \re^{-\ri\varepsilon_n T/\hbar} \}$ as its eigenvalues, and the expansion
coefficients of the approximate Floquet functions $u_n(x,0)$ as components of
its eigenvectors.         
           
Having computed the quasienergy spectrum for suitably chosen $n_{\rm max}$,
one faces a problem of visualization: If we were to plot all the $n_{\rm max}$ 
quasienergies obtained from the diagonalization, reduced to the fundamental 
Brillouin zone, there would be so many data points that one might not be able 
to recognize the important features, in particular so in cases where 
$n_{\rm max}$ is truly large. In addition, quasienergies resulting from 
high-lying states close to the ``truncation border'' might not be converged, 
and therefore should be left out. But it is not possible to identitify 
quasienergies of such ``high-lying'' Floquet states by their magnitude, 
because there exists no ``quasienergy ordering'' within the  Brillouin zone. 
A useful ordering scheme can be  established by considering the squared 
overlaps $| \langle \varphi_\ell | n \rangle |^2$  of the eigenvectors 
$\{ | n \rangle \}$ of the monodromy matrix with the basis states  
$\{ | \varphi_\ell \rangle \} $: If one assigns to each state $| n \rangle$
the index $\ell_{\rm max}(n)$ which maximizes that overlap, and then orders
the Floquet states and their quasienergies with respect to these indices, one
obtains at least a good pre-selection of the desired ``low-lying'' states.

\begin{figure}
\includegraphics[width = 0.9\linewidth]{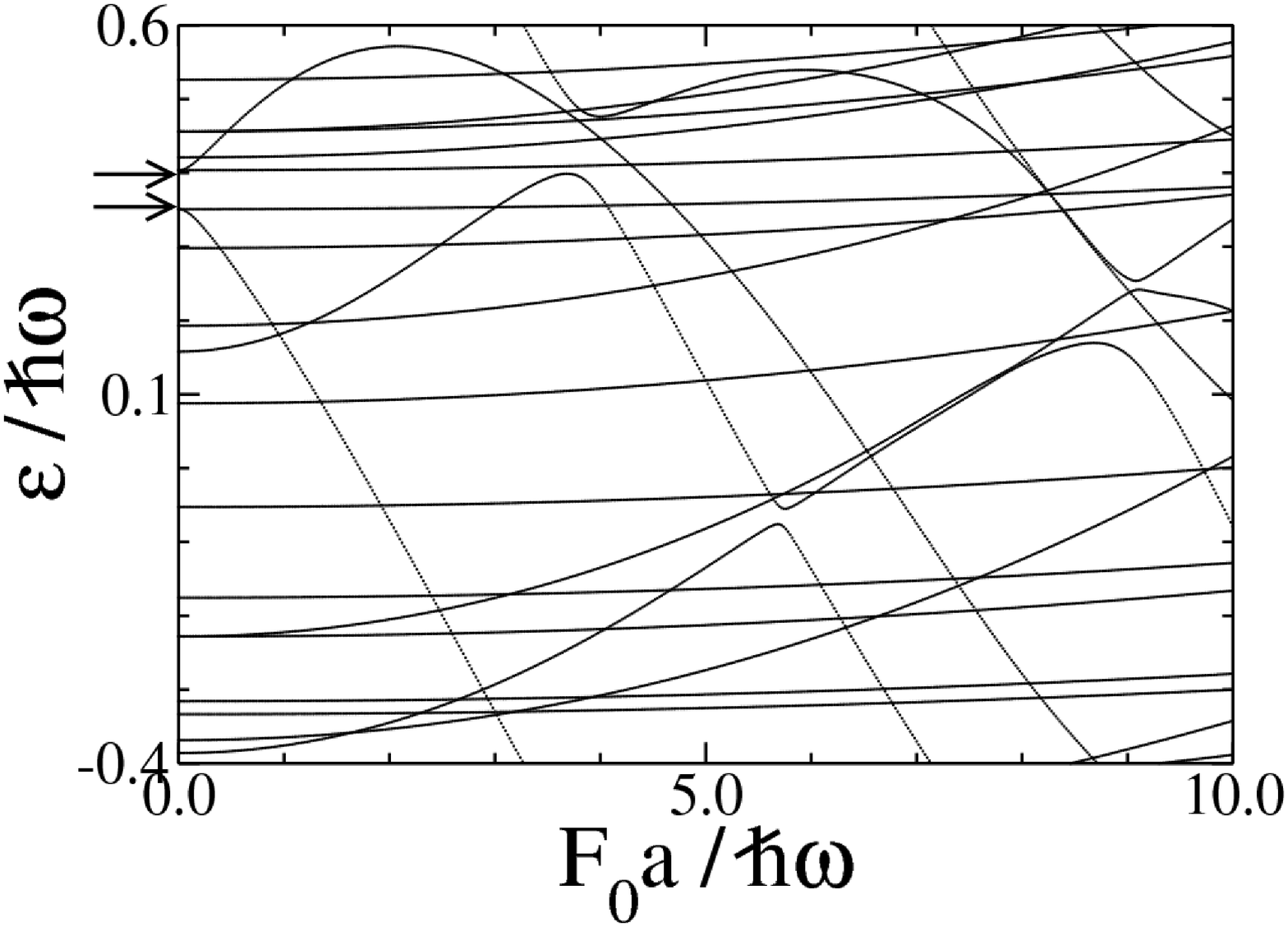}
\caption{One Brillouin zone of quasienergies for the ``driven particle in a 
	box'' with frequency~(\ref{eq:RLF}) vs.\ scaled driving amplitude;
	shown are the quasienergies originating from the 20 lowest box states. 
	Since the unperturbed box eigenstates $n = 1$ and $n = 2$ are almost
	resonant, the quasienergies emanating from these two states, indicated
	by the arrows, are well described by the RWA-expression~(\ref{eq:QTM}) 
	in the weak-driving regime $F_0a/(\hbar\omega) < 1$. The Floquet state 
	connected to the unperturbed ground state $n = 1$  exhibits an almost 
	linear, negative ac-Stark shift even under stronger driving, until it 
	undergoes at $F_0 a/(\hbar\omega) \approx 4.0$ an avoided crossing 
	with the Floquet state connected to the box state $n = 3$. This 
	anticrossing signals the presence of a $4$-photon resonance. The 
	further avoided crossing at $F_0 a/(\hbar\omega) \approx 5.7$ 
	indicates a $7$-photon resonance, involving the Floquet state emerging 
	from the box state $n = 4$.}
\label{F_3}	
\end{figure}

In Fig.~\ref{F_3} we display a quasienergy spectrum of the driven particle 
in the box~(\ref{eq:PIB}) which has been computed in this manner with 
$n_{\rm max} = 50$ for scaled driving strengths 
$0 \le F_0 a/(\hbar\omega) \le 10$, having plotted only quasienergies 
originating from the 20 lowest box states. The reader who is not already 
familiar with this kind of plot should dwell a moment to absorb its content: 
Because the driving frequency~(\ref{eq:RLF}) is slightly red-detuned from the 
transition between the box eigenstates $n = 1$ and $n = 2$, these two coupled
states constitute a two-level system for sufficiently low driving amplitudes. 
Thus, the quasienergies originating from these two states should conform 
to Eq.~(\ref{eq:QTM}) for weak driving, also assuming the validity of the 
rotating-wave approximation (RWA). Now the difference between red-detuning 
and blue-detuning shows up: According to Eq.~(\ref{eq:LI1}), in the case of 
a red-detuned driving frequency the quasienergy originating from the upper 
level is shifted upwards with increasing driving strength, whereas the 
quasienergy originating from the lower level is shifted downwards. In contrast, 
Eq.~(\ref{eq:LI2}) implies that in case of blue-detuning the quasienergy 
connected to the lower state would be shifted upwards, while the one connected 
to the higher state would be shifted downwards (see, {\em e.g.\/}, Fig.~1 in 
Ref.~\cite{LangemeyerHolthaus14}). Such shifts of the quasienergies against 
the unperturbed energy eigenvalues with increasing driving amplitude are 
generally known as {\em ac-Stark shifts\/}. Indeed, the numerical data shown
in Fig.~\ref{F_3} reveal precisely the expected pattern: The quasienergies 
emerging from the close-to-resonant unperturbed box eigenstates $n = 1$ and 
$n = 2$, indicated by the arrows in the left margin, are strongly affected 
even by weak driving, with the higher state $n = 2$ exhibiting a positive 
ac-Stark shift for $F_0a/(\hbar\omega) < 1$, whereas the lower state $n = 1$ is 
shifted downwards. For larger driving amplitudes the two-level-RWA naturally 
breaks down, and the ``higher'' quasienergy starts to bend, while the 
``lower'' quasienergy is shifted further downwards. If we now properly 
account for the Brillouin-zone structure~(\ref{eq:QEC}) of the quasienergy 
spectrum and label the quasienergy representatives such that 
$\varepsilon_n = \varepsilon_{(n,0)}$ reduces to the energy $E_n$ of the 
unpertubed $n$th box eigenstate in the limit of vanishing driving strength, 
the downward-shifted quasienergy representative connected to the unperturbed 
ground-state energy~$E_1$ in Fig.~\ref{F_3} carries the label $(1,0)$. When 
this representative reaches the lower border of the fundamental Brillouin zone,
the representative $(1,1)$ appears on its upper border, then undergoing a 
pronounced avoided crossing at $F_0 a/(\hbar\omega) \approx 4.0$ with a 
quasienergy representative labeled $(3,-3)$. It should be no surprise that 
the quasienergy connected to the ground-state energy of the box is anticrossed 
from {\em below\/}: Since there is no quasienergy ordering, the Floquet state 
originating from the box ground state no longer is a ground state. This 
anticrossing indicates a strong coupling between the two Floquet states 
involved, as corresponding to a {\em multi\-photon resonance\/}. 

The close interplay between ac-Stark shifts and multi\-photon resonances can 
be illustrated analytically with the help of the {\em linearly\/} polarized 
two-level system~(\ref{eq:TLL}): Starting from the transformation~(\ref{eq:TRA})   
one obtains the RWA-quasienergies~(\ref{eq:QTM}) as a first approximation, 
and then has to account for the counter-rotating terms. These terms have two 
effects: On the one hand, they cause a further shift of the quasienergies,
known as Bloch-Siegert shift. On the other, they couple the two Floquet states
when the shifted quasienergies attempt to cross at the boundary of the 
Brillouin zone, turning the Bloch-Siegert-shifted RWA crossings into avoided 
crossings (see Ref.~\cite{HolthausHone93} for a detailed elaboration of this 
program). Such couplings of two Floquet states, signaled by avoided crossings 
of their quasienergies, manifest themselves through strongly enhanced long-time
averaged transition probabilities under driving with constant 
amplitude~\cite{Shirley65}.

\begin{figure}
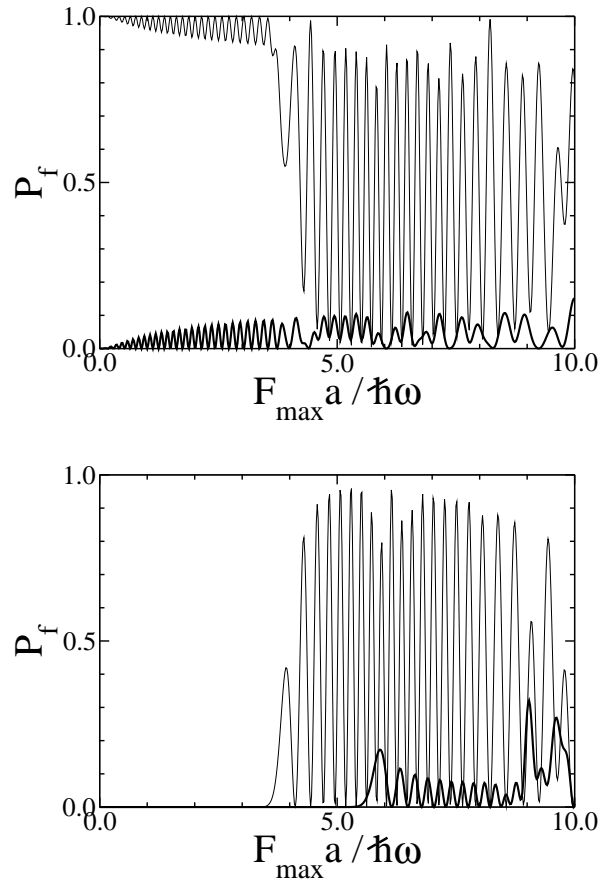

\includegraphics[width = 0.9\linewidth]{QES_Fig4a.eps}

\vspace*{4ex}

\includegraphics[width = 0.9\linewidth]{QES_Fig4b.eps}
\caption{Excitation of box eigenstates after Gaussian pulses~(\ref{eq:GEF})
	with width $\sigma/T = 10$ and near-resonant frequency~(\ref{eq:RLF}), 
	having started with the ground state $n = 1$ as initial state. The 
	upper panel shows the final transition probabilities~(\ref{eq:FTP}) 
	for $n = 1$ (thin line) and $n = 2$ (heavy line); the lower panel 
	those for $n = 3$ (thin line) and $n = 4$ (heavy line). Observe that 
	these latter states become populated once the pulses are so strong 
	that their envelopes reach the avoided crossings identified in 
	Fig.~\ref{F_3}.}    
\label{F_4}	
\end{figure}

These resonances constitute one of the most important building blocks for 
Floquet engineering, and will be put into active use in Chap.~\ref{SubSec46}. 
They also have profound consequences even when the external force is 
{\em not\/} perfectly periodic in time: If one initially populates the box 
ground state, say, and then subjects the system to a {\em pulse\/} with the 
frequency~(\ref{eq:RLF}), and with a slowly varying, smooth envelope~$F_0(t)$, 
the system's wave function tries to follow the instantaneous Floquet states 
in an adiabatic manner. However, when encountering an avoided crossing, 
Landau-Zener-type transitions to the anticrossing state 
occur~\cite{BreuerHolthaus89a,BreuerHolthaus89b,DreseHolthaus99}. In our 
example, such adiabatic following combined with Landau-Zener transitions among 
the anticrossing Floquet states will lead to a partial excitation of the second
excited box state $| \varphi_3 \rangle$ after a sufficiently strong pulse.  
From the difference $1 - (-3) = 4$ of the respective second entries into 
the label pairs $(1,1)$ and $(3,-3)$ of the participating quasienergy 
representatives one deduces that this transition corresponds to a $4$-photon 
resonance. In a similar manner, the avoided crossing visible in Fig.~\ref{F_3} 
at $F_0 a/(\hbar\omega) \approx 5.7$ indicates a $7$-photon resonance, 
involving quasienergy representatives $(1,1)$ and $(4,-6)$.

To confirm this quite general scenario, we choose a Gaussian envelope function
\begin{equation}
	F_0(t) = F_{\rm max}\exp\left(-\frac{t^2}{2\sigma^2}\right) \; ,
\label{eq:GEF}	
\end{equation}
start with the box ground state $\varphi_1(x) = \psi(x,-\infty)$ as initial
state, and compute the wave function $\psi(x,+\infty)$ after the pulse. In 
Fig.~\ref{F_4} we
show the transition probabilities
\begin{equation}
	P_f(n) = \big| \langle \varphi_n | \psi(+\infty) \rangle \big|^2
\label{eq:FTP}
\end{equation}
effectuated by such pulses with a width of $\sigma/T = 10$, as functions of
the scaled maximum amplitude. As expected, the second excited box eigenstate
$n = 3$ is signficantly populated once the maximum scaled amplitude reaches 
the large avoided crossing showing up in Fig.~\ref{F_3} at 
$F_0 a/(\hbar\omega) \approx 4.0$; the third excited state $n = 4$ appears 
when the amplitude even reaches the further avoided crossing at
$F_0 a/(\hbar\omega) \approx 5.7$. The oscillating excitation patterns are
due to the fact that the slowly varying envelope traverses the anticrossings
twice, during its rise and during its subsequent decrease, so that the
two Floquet states which have been populated after the first traversal 
interfere at the second. This leads to Stueckelberg oscillations, or Ramsey 
interference fringes. We remark that such oscillations have been detected
in experiments with potassium~\cite{BaruchGallagher92} and 
helium~\cite{YoakumEtAl92} Rydberg atoms driven by short microwave pulses.

\subsection{The extended Hilbert space}
\label{SubSec34}
  
Coming back to the attempt~(\ref{eq:NAT}) to define an operator~$G$ on
${\mathcal H}$ by wilfully selecting one particular representative 
$\varepsilon_n$ from each class~(\ref{eq:QEC}), we are faced with a dilemma: 
Which representatives should one pick out in the example depicted in 
Fig.~\ref{F_3}? Those representatives which connect to the unperturbed box 
energies in the limit of vanishing driving strength? But then the chosen 
$\varepsilon_n$ would not be those representatives which undergo the avoided 
crossings, while these mark the all-important multi\-photon resonances. Or 
should one take representatives coupled at an avoided crossing? Apart from 
the fact that this would lead to assignment problems when multiple avoided 
crossings appear, such representatives would not end up at the unperturbed 
energy eigenvalues, which appears strange in the perturbative regime of 
small driving amplitudes. Thus, with the exception of sufficiently simple
systems for which there exists a ``canonical'' choice of quasienergy
respresentatives, it seems advisable to abandon the attempt~(\ref{eq:NAT}) 
altogether and to formulate the theory in an invariant manner, such that 
unnatural distinctions of individual quasienergy representatives are not made. 
Is this possible?

It is. By inserting the Floquet states~(\ref{eq:FST}) into the Schr\"odinger
equation~(\ref{eq:SGL}) one easily confirms that the Floquet functions
$| u_n(t) \rangle $ satisfy the identity
\begin{equation}
	K | u_n(t) \rangle = \varepsilon_n | u_n(t) \rangle \; ,
\label{eq:EEH}
\end{equation}
where we have introduced the quasienergy operator
\begin{equation}
	K = H(t) - \ri\hbar\frac{\rd}{\rd t} \; .
\label{eq:EOK}
\end{equation}
If at this point the time~$t$ no longer is regarded as the evolution 
variable, but rather as a {\em coordinate\/} on the same footing as~$x$, 
this Eq.~(\ref{eq:EEH}) becomes an eigenvalue equation in an {\em extended 
Hilbert space\/} of $T$-periodic functions, denoted 
$L_2[0,T] \otimes {\mathcal H}$, where ${\mathcal H}$ is the space that 
$H(t)$ acts on, as before. Now, taking one particular $T$-periodic solution 
$|u_n(t) \rangle$ to Eq.~(\ref{eq:EEH}) with eigenvalue $\varepsilon_n$,
and multiplying by $\re^{\ri m \omega t}$, where $\omega$ is given by
Eq.~(\ref{eq:DFO}) and $m = 0,\pm 1,\pm2,\ldots$, the product
$|u_n(t) \rangle \re^{\ri m \omega t}$ again is $T$-periodic, obeying  
\begin{equation}
	K | u_n(t) \rangle \re^{\ri m \omega t}
	= (\varepsilon_n + m \hbar \omega) 
	| u_n(t) \rangle \re^{\ri m \omega t} \; .
\end{equation}
That is, all the quasienergy representatives of a given class~(\ref{eq:QEC}) 
appear as individual solutions to this eigenvalue equation~(\ref{eq:EEH}) 
in $L_2[0,T] \otimes {\mathcal H}$. But they all lead to the {\em same\/}
Floquet state~(\ref{eq:FST}) in ${\mathcal H}$, since 
\begin{eqnarray}
	& & 
	\Big(| u_n(t) \rangle \re^{\ri m \omega t} \Big) 
	\exp\big( -\ri[\varepsilon_n + m\hbar\omega]t/\hbar \big)
\nonumber \\	& = &	
	| u_n(t) \rangle \re^{-\ri \varepsilon_n t/\hbar} \; ,
\end{eqnarray}	
so that the ``$\bmod \; \hbar\omega$-indeterminacy'' drops out.

The extended Hilbert space appears to have been introduced into the physics 
literature by Hideo Sambe~\cite{Sambe73}; it plays a major role in the 
rigorous mathematical analysis of periodically time-dependent quantum 
systems~\cite{Howland79,Howland89,Howland92b,Joye94}. To fully appreciate 
what is going on here it is helpful to make another digression and to consider 
a classical system defined by some explicitly time-dependent Hamiltonian 
function $H_{\rm cl}(p,x,t)$, giving rise to the Hamiltonian equations   
\begin{eqnarray}
	\frac{\rd x}{\rd t} & = & 
	\phantom{-}\frac{\partial H_{\rm cl}}{\partial p} \; ,
\nonumber	\\
	\frac{\rd p}{\rd t} & = & -\frac{\partial H_{\rm cl}}{\partial x} \; .
\label{eq:OHS}
\end{eqnarray}		
If one wishes to treat this time-dependent system in analogy to an autonomous 
one, one regards the time~$t$ as a coordinate, which then naturally possesses 
a canonically conjugate momentum variable~$p_t$, leading to an extended
phase space $\big\{ (p,p_t,x,t) \big\}$. Next, taking the function
\begin{equation}
	K_{\rm cl}(p,p_t,x,t) = H_{\rm cl}(p,x,t) + p_t
\label{eq:AHF}	
\end{equation}	
as a Hamiltonian in this extended phase space, one needs an evolution variable
which parametrizes the flow it generates; let us denote this evolution variable
by~$\tau$. One then is led to the augmented Hamiltonian system
\begin{eqnarray}
	\frac{\rd x}{\rd \tau} & = &
	\phantom{-}\frac{\partial K_{\rm cl}}{\partial p}
	\; = \; \phantom{-}\frac{\partial H_{\rm cl}}{\partial p} \; ,
\nonumber	\\
	\frac{\rd t}{\rd \tau} &  = & 
	\phantom{-}\frac{\partial K_{\rm cl}}{\partial p_t}  
	\; = \quad 1 \; ,
\nonumber	\\
	\frac{\rd p}{\rd \tau} & = &
	-\frac{\partial K_{\rm cl}}{\partial x}
	\; = \; -\frac{\partial H_{\rm cl}}{\partial x} \; ,
\nonumber	\\	
	\frac{\rd p_t}{\rd \tau} & = &
	-\frac{\partial K_{\rm cl}}{\partial t} \; ,
\label{eq:EHE}
\end{eqnarray}	
revealing the close connection of this extended, but autonomous problem to 
the original one: The second of these Eqs.~(\ref{eq:EHE}) tells us that the 
artificial evolution variable~$\tau$ coincides with~$t$, up to an irrelevant 
constant which may be taken to be zero. The first and the third equation then 
reproduce the original system~(\ref{eq:OHS}), while the fourth one makes sure 
that the ``Kamiltonian''~(\ref{eq:AHF}) is conserved, as it should, because it
does not depend on the time~$\tau$ --- recall that $t$ has been promoted to
a coordinate! The use of these extended-phase space techniques allows one, 
among others, to adapt the semiclassical Einstein-Brillouin-Keller quantization
rules for energy eigenstates such that they lend themselves to a semiclassical 
determination of Floquet states and their quasienergies~\cite{BreuerHolthaus91}.     		

Obviously the quasienergy operator~(\ref{eq:EOK}), acting on the extended 
Hilbert space $L_2[0,T] \otimes {\mathcal H}$, is precisely the quantum 
counterpart of the augmented Hamiltonian function~(\ref{eq:AHF}) on the 
extended phase space, with the classical ``time-momentum'' $p_t$ having been 
replaced by the momentum operator ({\em sic!}\/)
\begin{equation}
	p_t = \frac{\hbar}{\ri} \frac{\rd}{\rd t} \; ; 
\label{eq:MOO}
\end{equation}	
observe that the periodic boundary condition in~$t$ makes sure that 
this operator is Hermitian. The noteworthy fact that this momentum 
operator~(\ref{eq:MOO}) enters only {\em linearly\/} into the quasienergy 
operator~$K$ is responsible for the fact that the spectrum of~$K$, consisting 
of all representatives of all quasienergies, is unbounded from below; a 
property which appears quite unusual in nonrelativistic quantum mechanics. 
Notwithstanding this technically difficult feature, the eigenvalue 
equation~(\ref{eq:EEH}) now plays the role of a {\em stationary\/} 
Schr\"odinger equation~\cite{Sambe73}. 

This recognition leads us to a further element of the Floquet picture: What, 
then, would be the analog of the time-dependent Schr\"odinger equation in 
the extended Hilbert space? The answer to this question is suggested by the 
classical construction~(\ref{eq:EHE}): With $t$ being a coordinate, we require 
a new evolution variable~$\tau$, and thus consider states $| \Psi(\tau,t) \db$, 
where, following Sambe~\cite{Sambe73}, the double bracket symbol is meant to 
indicate that these states reside in $L_2[0,T] \otimes {\mathcal H}$, rather 
than in ${\mathcal H}$. For consistency, these states then should obey the 
Schr\"odinger-like evolution equation
\begin{equation}
	\ri\hbar\frac{\rd}{\rd \tau} | \Psi(\tau,t) \db
	= K | \Psi(\tau,t) \db \; .
\label{eq:ESE}
\end{equation}       
Moreover, in analogy to the classical procedure the actual physical state 
$| \psi(t) \rangle$ in ${\mathcal H}$ should be recovered from
$| \Psi(\tau,t) \db$ by equating $\tau$ and $t$:
\begin{equation}
	| \psi(t) \rangle = | \Psi(\tau,t) \db \Big|_{\tau = t} \; . 
\label{eq:PRO}
\end{equation}
Indeed, this requirement gives
\begin{eqnarray}
	\ri\hbar\frac{\rd}{\rd t}    | \psi (t) \rangle & = &
	\ri\hbar\frac{\rd}{\rd \tau} | \Psi (\tau,t) \db \Big|_{\tau = t} +
	\ri\hbar\frac{\rd}{\rd t}    | \Psi (\tau,t) \db \Big|_{\tau = t}
\nonumber \\	& = &
        \left( H(t) - \ri\hbar\frac{\rd}{\rd t} \right)
	| \Psi (\tau,t) \db \Big|_{\tau = t} 
\nonumber \\	& & 	
	+ \ri\hbar\frac{\rd}{\rd t} | \Psi (\tau,t) \db \Big|_{\tau = t}	
\nonumber \\	& = &
	H(t) | \psi(t) \rangle \; ,	
	\phantom{\frac{\rd}{\rd t}}
\end{eqnarray}
where Eq.~(\ref{eq:ESE}) has been used in the second step, together with the
definition~(\ref{eq:EOK}) of the quasienergy operator~$K$: The evolution
equation~(\ref{eq:ESE}) in $L_2[0,T] \otimes {\mathcal H}$, together with
the prescription~(\ref{eq:PRO}) for ``projecting back'' from  
$L_2[0,T] \otimes {\mathcal H}$ to the physical Hilbert space ${\mathcal H}$,
implies the correct Schr\"odinger equation~(\ref{eq:SGL}). This suggests that 
the states $|\psi(t)\rangle$ of a periodically time-dependent quantum system 
may be regarded as ``projection shadows'', in the sense of Eq.~(\ref{eq:PRO}),
of larger objects which are governed by a ``Kamiltonian'' involving a momentum 
operator only to its first power. This is what makes these systems so exciting,
both literally and metaphorically speaking, and this is what underlies much of 
their potential for quantum engineering.

The evolution equation~(\ref{eq:ESE}) provides the key for establishing 
an adiabatic principle for Floquet states~\cite{BreuerHolthaus89a,
BreuerHolthaus89b,DreseHolthaus99,EckardtHolthaus08}, for connecting multiphoton
transitions, as exemplified in Fig.~\ref{F_4}, to Landau-Zener transitions among
Floquet states~\cite{BreuerHolthaus89a,BreuerHolthaus89b,DreseHolthaus99}, and 
for superadiabatic Floquet analysis~\cite{DreseHolthaus99}. But is there also 
something more tangible, beyond formal consistency? 

There is. The traditionally trained reader may have been severely worried by 
Fig.~\ref{F_4}: The unperturbed box Hamiltonian~(\ref{eq:UPO}) is invariant 
under the reflection $x \to -x$, so that its eigenfunctions $\varphi_n(x)$ 
possess the definite parity $(-1)^{n+1}$. Moreover, the dipole operator~$x$ 
which mediates the time-periodic perturbation according to Eq.~(\ref{eq:PIB}) 
only connects states with different parity, meaning that all dipole matrix
elements~(\ref{eq:DME}) between states of equal parity vanish. Hence, the 
transition from $n = 1$ to $n = 2$ is ``dipole-allowed'', whereas the 
transition from $n = 1$ to $n = 3$ is ``dipole-forbidden'' --- in apparent 
contradiction to the numerical results shown in Fig.~\ref{F_4}. Even more, 
the occupation of the seemingly favored final state $n = 2$ would actually 
be adiabatically suppressed if the pulses were made longer. What is happening 
here?

The answer to this pertinent question, of course, lies in the observation that 
the reflection symmetry only refers to the unperturbed system~(\ref{eq:UPO}), 
and therefore remains meaningful as long as the driving force remains 
``perturbatively weak''. But if this is no longer the case, because the 
dimensionless quantity $F_0 a/(\hbar\omega)$ is taken to be on the order of 
unity or larger, then it is not the symmetry of $H_0$ in ${\mathcal H}$ 
which matters, but rather that of $K$ in $L_2[0,T] \otimes {\mathcal H}$. 
Now the quasienergy operator~(\ref{eq:EOK}) constructed from the full 
Hamiltonian~(\ref{eq:PIB}) remains invariant under the spatio-temporal 
transformation
\begin{equation}   
	P : \left\{ \begin{array}{ccl}
		x & \to & -x	\\
		t & \to & t + T/2	\; ;
		    \end{array} \right.	
\end{equation}
since $P^2 = {\rm id}$ (keeping in mind that $t + T$ equals $t$ in 
$L_2[0,T] \otimes {\mathcal H}$) the eigenfunctions of $K$ acquire a definite
sign under this generalized parity operation. Assuming the same labeling as
before, so that the quasienergy representative $(n,0)$ connects to the
unperturbed box energy $E_n$ in the limit of vanishing driving strength, one 
finds that a representative labeled $(n,m)$ is associated with the 
generalized parity $(-1)^{n+m+1}$ --- for a proof, consider the particular
parameter $F_0 a/(\hbar\omega) = 0$, for which the Floquet functions coincide 
with the unperturbed box eigenfunctions, times $\re^{\ri m \omega t}$. Now 
there is the famous von Neumann-Wigner noncrossing rule: Eigenvalues of a 
Hermitian operator do not cross if only one single parameter is varied, unless 
there is a symmetry which allows such crossings~\cite{NeumannWigner29}. 
Applied to the quasienergy spectrum of the periodically driven particle 
in the box, as displayed in Fig.~\ref{F_3}, this means that quasienergy 
representatives possessing the same generalized parity are not allowed to 
cross upon variation of $F_0 a/(\hbar\omega)$, and therefore produce 
multiphoton resonances, whereas representatives having different generalized 
parities do not ``feel'' each other and may cross. Indeed, the representatives 
$(1,1)$ and $(3,-3)$ repelling each other and thereby effectuating the 
$4$-photon-resonance located at $F_0 a/(\hbar\omega) \approx 4.0$ both 
fall into the same generalized parity class, as do the representatives 
$(1,1)$ and $(4,-6)$ which form the $7$-photon resonance at 
$F_0 a/(\hbar\omega) \approx 5.7$. Thus, the concept of the extended Hilbert 
space is essential for understanding nonperturbative excitation patterns.

\subsection{Coarse graining}
\label{SubSec35}
				
\begin{figure}
\includegraphics[width = 0.9\linewidth]{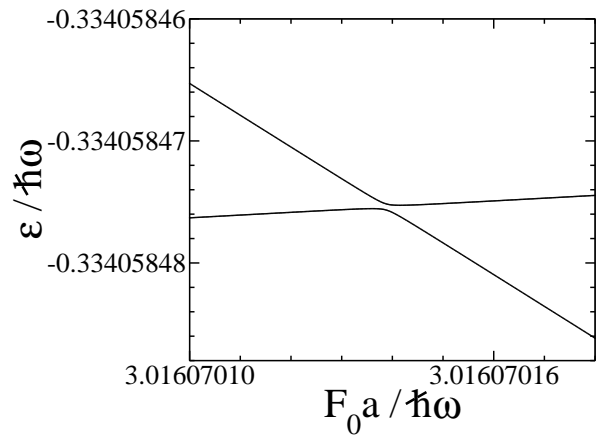}
\caption{Magnified detail of Fig.~\ref{F_3}, showing an anticrossing between
	quasienergy representatives $(1,0)$ and $(6,-13)$. Almost needless to 
	stress: Observe the scales!}
\label{F_5}	
\end{figure}

Taking the noncrossing rule seriously, about half of the apparent crossings 
observed in Fig.~\ref{F_3} actually should be non-resolved anticrossings.
This is exemplified by Fig.~\ref{F_5}, which shows a magnification of a
detail far below the resolution of Fig.~\ref{F_3}, namely, the expected
anticrossing of the quasienergy representatives $(1,0)$ and $(6,-13)$, 
corresponding to an extremely narrow $13$-photon resonance. In a similar 
manner one confirms the validity of the noncrossing rule at other instances,
although it becomes hard to maintain the required numerical accuracy at
high-order resonances.

But now we have opened Pandora's box. Recall that each of the infinitely many
box eigenstates plants one quasienergy representative into the fundamental
Brillouin zone, so that, at least for a generic choice of the driving 
frequency, the quasienergy spectrum will be ``dense'' already for
$F_0 a/(\hbar\omega) = 0$. If the previously constructed picture, resulting
from finite-subspace approximations, actually still holds water even when
considering the full infinite-dimensional system, this would imply infinitely
many anticrossings ``on arbitrarily small scales'' within an arbitrarily small 
interval of the scaled driving strength; hence, it seems unlikely that the 
``true'' quasienergies are differentiable. There are resonances virtually 
everywhere --- although the vast majority of them evidently have to be of
{\em very\/} high order, such that they should not matter much. 

At this point, the attitudes of mathematicians and physicists may differ. The 
mathematician will observe that, when studying systems like the driven particle
in the box~(\ref{eq:PIB}), one is dealing with the problem of the perturbation 
of a dense point spectrum, which is far from trivial. In particular, one may 
pose the question whether there exists a transition from a dense pure point 
spectrum to an absolutely continuous one at a certain nonzero driving 
strength~\cite{Howland92}. While a dense point quasienergy spectrum gives 
rise to quasiperiodic motion~\cite{HoggHuberman82}, a continuous quasienergy 
spectrum would lead to diffusive energy growth~\cite{BunimovichEtAl91}; the 
question under what conditions there can be a transition from one type of 
motion to the other constitutes the content of the so-called {\em quantum 
stability problem\/}~\cite{Howland92,Combescure88}. This problem has been 
studied in great depth for the model of the ``$\delta$-kicked 
rotator''~\cite{CasatiEtAl79}: While its quasienergy spectrum has a continuous 
component when the frequency of the kicks is rationally related to the 
frequencies of the free rotator~\cite{CasatiGuarneri84}, it is pure point 
otherwise~\cite{Howland92}. Eminent contributions to the general subject 
of quantum stability have been made by James Howland, who has established 
conditions guaranteeing the absence of an absolutely continuous 
spectrum~\cite{Howland92,Howland89}. Using these, it has been shown that 
linearly driven anharmonic oscillators with a superquadratic potential, 
such as the driven particle in a box, are ``stable'' in this 
sense~\cite{Howland92b}.

The physicist, on the other hand, may feel that a mathe\-matical model like 
that given by Eq.~(\ref{eq:PIB}) provides only a limited description of some 
larger, experimentally accessible system, so that it might not always be 
meaningful to solve such a restricted model on a level at which it no longer 
applies. To be definite, a tiny avoided quasienergy crossing of width 
$\delta\varepsilon$, such as shown in Fig.~\ref{F_5}, is associated with a 
large time scale on the order of $\hbar/\delta\varepsilon$, and would not be 
detectable in shorter measurements. In experiments with pulses such tiny 
anticrossings would be traversed in a perfectly diabatic manner and therefore 
remain hidden, {\em unless\/} the time scale characterizing the change of the 
pulses' envelope matches $\hbar/\delta\varepsilon$. But then, who could design 
pulses with ``infinitely slowly'' varying envelopes?

Thus, in many cases of experimental interest it may be reasonable to ignore
infinitely many quasienergy anti\-crossings below a certain scale determined
by the respective set-up. This ``coarse graining'' strategy had automatically
been implied by Fig.~\ref{F_3}: The $20$ ``lowest'' quasienergies displayed
there actually undergo avoided crossings with ``higher'' states which have
been omitted, but these anticrossings are too small to be seen, so that the
remaining lines appear smooth on the scale of Fig~\ref{F_3}. That same coarse
graining approach, as well as its limitations, will be met again in the 
following calculations of quasienergy bands for periodically driven optical
lattices. If coarse graining is feasible, one may construct an effective
time-independent Hamiltonian which describes the dynamics within some
restricted subspace~\cite{GoldmanDalibard14}; if not, efficient coupling among 
Floquet states causes heating of the driven system. 

On a technical level, coarse graining amounts to constructing an integrable
approximation to a near-integrable system. This usually involves some sort 
of high-frequency approximation for averaging out residual oscillating 
perturbations, similar to the rotating-wave approximation to the linearly 
polarized two-level system~(\ref{eq:TLL}). A systematic discussion of such 
approximation schemes within the Floquet picture has recently been given in 
Ref.~\cite{EckardtAnisimovas15}. It might also be worthwhile to point out 
that the quantum stability problem still remains virtually unexplored for 
periodically forced systems comprising many interacting particles: In view 
of the tremendous density of states resulting from folding their unperturbed
energy spectra into the Brillouin zone, it might be hard to find generic 
driven many-body systems that are truly stable in the sense of 
Refs.~\cite{Howland92,Combescure88}.       
 
For completeness, we briefly mention another important branch of the theory:
When considering, for instance, a hydrogen atom exposed to a monochromatic
classical driving force, one is dealing with the perturbation of eigenstates
embedded in a continuous spectrum. The resulting Floquet states then are 
{\em resonances\/} with a finite lifetime, as described mathematically by
complex poles of the resolvent of the quasienergy operator~\cite{Yajima82}.   
This is exploited in practical computations of, {\em e.g.\/}, ionization
probabilities with the help of complex scaling techniques~\cite{ChuTelnov04}, 
but these are not required for the present purposes.

\section{Floquet engineering with optical lattices}
\label{S_4}

We now turn to the central topic of this tutorial, aiming at the deliberate
manipulation, and control, of quantum dynamics in driven optical lattices. 
Thus, we consider a quantum particle which moves in a spatially periodic 
potential~$V$ with ``lattice constant''~$a$,  
\begin{equation}
	V(x) = V(x + a) \; ,
\label{eq:PPV}	
\end{equation}	
while it is acted on by a spatially homogeneous, temporally periodic force~$F$ 
with period~$T$,
\begin{equation}
	F(t) = F(t+T) \; .
\label{eq:PKF}	 
\end{equation}	
Assuming dipole coupling, as for the particle in the box~(\ref{eq:PIB}), 
the Hamiltonian takes the form
\begin{equation}
	\widetilde{H}(x,t) = -\frac{\hbar^2}{2M} \frac{\rd^2}{\rd x^2}
	+ V(x) - x F(t) \; ,
\label{eq:HDC}	
\end{equation}
and the task again is to characterize {\em all\/} solutions to the
time-dependent Schr\"odinger equation
\begin{equation}
	\ri\hbar\frac{\rd}{\rd t} \widetilde{\psi}(x,t) =
	\widetilde{H}(x,t) \widetilde{\psi}(x,t) \; .
\end{equation}
We proceed in three steps: In Chap.~\ref{SubSec41} we provide a general
formulation of our approach, introducing quasienergy bands and quasienergy
dispersion relations. In Chap.~\ref{SubSec42} we then discuss driven optical
cosine lattices, and show how their quasienergy dispersion relations are 
computed in practice, combining the essentials of Secs.~\ref{S_2} and 
\ref{S_3}. In the remaining Chaps.~\ref{SubSec43} to \ref{SubSec46} we then 
explain specific examples of Floquet engineering,  showing how to create 
quasienergy dispersion relations with certain desired properties. Throughout, 
we will make heavy use of the theoretical concepts introduced in the previous 
Sec.~\ref{S_3} --- the quasienergy Brillouin zone, ac-Stark shifts, multiphoton
resonances, the extended Hilbert space, and coarse graining ---, and 
demonstrate their experimental implications.

\subsection{Spatio-temporal Bloch waves}
\label{SubSec41}

In order to exploit both the spatial periodicity~(\ref{eq:PPV}) and the 
temporal periodicity~(\ref{eq:PKF}), we perform the unitary transformation   
\begin{equation}
	\widetilde{\psi}(x,t) = 
	\exp\left( \frac{\ri}{\hbar} x \int_0^t \rd \tau \, F(\tau) \right)
	\psi(x,t) 
\label{eq:KHT}
\end{equation}
which brings the coupling of the external force to the particle into a 
different, but quite familiar guise: Observing 
\begin{eqnarray}
	\ri\hbar\frac{\rd}{\rd t} \widetilde{\psi}(x,t) & = & 	
	\exp\left( \frac{\ri}{\hbar} x \int_0^t \rd \tau \, F(\tau) \right)
\nonumber \\	& & \times 	
	\left( \ri\hbar\frac{\rd}{\rd t} \psi(x,t) - x F(t) \psi(x,t) \right)
\end{eqnarray}	 	
and invoking the identity~(\ref{eq:BCH}), we are led to the transformed
Schr\"odinger equation
\begin{equation}
	\ri\hbar\frac{\rd}{\rd t} \psi(x,t) = H(x,t) \psi(x,t)
\label{eq:TSE}
\end{equation}	
with the new Hamiltonian
\begin{equation}
	H(x,t) = 
	\frac{1}{2M}\left( p + \int_0^t \! \rd \tau \, F(\tau) \right)^2
	+ V(x) \; .
\label{eq:HXT}
\end{equation}
Here the homogeneous force couples to the particle in the same gauge-invariant
manner as a homogeneous electric field would couple to a particle with 
charge~$e$: Introducing the space-independent ``vector potential''
\begin{equation}
	e A(t) = -\int_0^t \! \rd \tau \, F(\tau) \; ,
\end{equation}	  
so that
\begin{equation}
	F(t) = -e \frac{\rd A(t)}{\rd t} \; ,
\end{equation}
one has 
\begin{equation}
	H(x,t) = 
	\frac{1}{2M}\big( p - e A(t) \big)^2 + V(x) \; .
\end{equation}
But the actual reason for resorting to the transformation~(\ref{eq:KHT}) 
lies elsewhere: If we now stipulate that the force does not contain a dc 
component, so that
\begin{equation}
	\frac{1}{T}\int_0^T \! \rd t \, F(t) = 0 \; ,
\label{eq:ZAC}	
\end{equation}
this Hamiltonian~(\ref{eq:HXT}) is periodic in both space and time:
\begin{equation}
	H(x,t) = H(x+a,t) = H(x,t+T) \; .
\label{eq:HST}	
\end{equation}
Accordingly, we can apply both the common Bloch theory as known from 
solid-state physics~\cite{AshcroftMermin76,Harrison89}, and the Floquet 
theory outlined in Sec.~\ref{S_3}. Hence, a Hamiltonian with the two-fold 
translational symmetry~(\ref{eq:HST}) gives rise to a complete set of 
spatio-temporal Bloch waves~\cite{Holthaus92,DreseHolthaus97b,
ArlinghausHolthaus11}       
\begin{equation}
	\psi_{n,k}(x,t) = 
	\exp\big[ \ri k x - \ri \varepsilon_n(k) t/\hbar\big]
	u_{n,k}(x,t)
\label{eq:STB}
\end{equation}
characterized simultaneously by a wave number~$k$ and a 
quasienergy~$\varepsilon_n(k)$; as in Sec.~\ref{S_2}, an additional index~$n$ 
is required for distinguishing different {\em quasienergy bands\/}. The 
Bloch-Floquet functions $u_{n,k}(x,t)$ now embody both translational 
symmetries, so that
\begin{equation}
	u_{n,k}(x,t) = u_{n,k}(x+a,t) = u_{n,k}(x,t+T) \; .
\end{equation}
The rationale behind the introduction of these quasi\-energy bands lies in the
fact that they open up far-reaching analogies: A particle in the $n$th energy 
band $E_n(k)$ of a lattice $V(x) = V(x + a)$ {\em without\/} driving force 
$F(t)$ is described by a wave packet  
\begin{eqnarray}
	\psi_n(x,t) & = & \sqrt{\frac{a}{2\pi}} \int \! \rd k \, g_n(k)
\nonumber \\	& & \times	 
	\exp[\ri k x - \ri E_n(k)t/\hbar] u_{n,k}(x)
	\phantom{\sum} 
\end{eqnarray}
built from the usual Bloch waves~(\ref{eq:BLA}), where the integration
ranges over one Brillouin zone $-\pi/a \leq k < \pi/a$ in $k$-space. 
Assuming the $k$-space distribution $g_n(k)$ to be sufficiently smooth,
and well centered around its first moment    
\begin{equation}
	k_c =  \int \! \rd k \, k | g_n(k) |^2 \; ,
\label{eq:MOM}	
\end{equation}	  
the packet's group velocity is given by
\begin{equation}
	v_g = \frac{1}{\hbar} \left. \frac{\rd E_n(k)}{\rd k} \right|_{k_c}
	\; .
\end{equation}
Moreover, if an arbitrary, but weak homogeneous force $f(t)$ acts on the 
system, such that it does not produce appreciable interband transitions, 
the moment~(\ref{eq:MOM}) evolves in time according to the semiclassical 
``acceleration theorem''~\cite{AshcroftMermin76}	
\begin{equation}
	\hbar \frac{\rd k_c(t)}{\rd t} = f(t) \; .
\label{eq:ACC}
\end{equation}	 
In complete analogy, a particle prepared in the $n$th quasienergy band of a 
periodically driven lattice is given by a packet~\cite{ArlinghausHolthaus11} 
\begin{eqnarray}
	\psi_n(x,t) & = & \sqrt{\frac{a}{2\pi}} \int \! \rd k \, g_n(k)
\nonumber \\	& & \times	 
	\exp[\ri k x - \ri\varepsilon_n(k)t/\hbar] u_{n,k}(x,t)
	\phantom{\sum} 
\label{eq:WPA}
\end{eqnarray}
made up from the spatio-temporal Bloch waves~(\ref{eq:STB}), with the 
occupation amplitudes $g_n(k)$ remaining {\em constant\/} in time, despite the 
action of the external force. This is an immediate consequence of Assertion~3 
formulated in Chap.~\ref{SubSec31}: While the driving force may give rise to 
even violent transitions between the unperturbed energy bands, these transitions
are incorporated into the spatio-temporal Bloch basis~(\ref{eq:STB}), to the 
effect that the occupation numbers of these time-dependent basis states are 
preserved in time. In addition, assuming the quasienergy dispersion relation 
$\varepsilon_n(k)$ to be sufficiently smooth,\footnote{The proposition of
	sufficient smoothness may not always be satisfied; it requires 
	that coarse graining works reasonably well: See the quasienergy 
	bands displayed in Chaps.~\ref{SubSec43} to \ref{SubSec46}~!}   
the cycle-averaged group velocity of such a packet~(\ref{eq:WPA}) is 
determined by its derivative,
\begin{equation}
	\overline{v}_g = \frac{1}{\hbar} 
	\left. \frac{\rd \varepsilon_n(k)}{\rd k} \right|_{k_c} \; .
\end{equation}	
Finally, if the total driving force $F(t) + f(t)$ consists of a strong 
time-periodic component~$F(t)$ which creates the Floquet states, and an 
additional weak component~$f(t)$, one again finds an acceleration theorem of 
the form~(\ref{eq:ACC})~\cite{ArlinghausHolthaus11}. Thus, the introduction
of quasienergy dispersion relations $\varepsilon_n(k)$ allows one to assess
the essential dynamics in homogeneously driven lattices in close analogy
to the approach taken in energy band-based solid state physics.      

Inserting the ansatz~(\ref{eq:STB}) into the Schr\"odinger 
equation~(\ref{eq:TSE}), and once again using Eq.~(\ref{eq:BCH}), we obtain
the quasienergy eigenvalue equation which determines the quasienergy band
structure:
\begin{widetext}
\begin{equation}
	\left[ \frac{1}{2M}
	\left( p + \hbar k + \int_0^t \! \rd \tau \, F(\tau) \right)^2	
	+ V(x) - \ri\hbar\frac{\rd}{\rd t} \right]
	u_{n,k}(x,t) = \varepsilon_n(k) u_{n,k}(x,t) \; .
\label{eq:QBS}
\end{equation}	
\end{widetext}
This is the concrete form of the eigenvalue equation~(\ref{eq:EEH}) required
here, defined on the extended Hilbert space $L_2[0,T] \otimes {\mathcal H}$ 
which hosts the eigenfunctions $u_{n,k}(x,t)$. Viewed from the perspective of 
this extended Hilbert space, which prompts us, in the sense discussed in 
Chap.~\ref{SubSec34}, to treat the time variable~$t$ as a coordinate on the 
same footing as~$x$, a time-periodically driven lattice thus appears as a 
``spatio-temporal lattice'', with the drive extending the spatial periodicity 
imposed by the lattice potential~$V$ to the additional ``time coordinate''. 
Even if the spatial lattice potential~$V$ is kept fixed, one can manipulate the 
quasienergy dispersion relations~$\varepsilon_n(k)$ of this spatio-temporal 
lattice by suitably adjusting its temporal component, {\em i.e.\/}, the 
parameters of the driving force. This sets the scope for Floquet engineering 
of dispersion relations, and opens up a wide field of possibilities far beyond 
the reach of customary solid-state physics. 

In passing, we remark that the ``zero average'' condition~(\ref{eq:ZAC}) 
imposed on the force $F(t)$ can still be relaxed: Quasienergy bands exist 
when~\cite{ArlinghausHolthaus11}
\begin{equation}
  	\int_0^T \! \rd t \, F(t) = r \times \hbar \frac{2\pi}{a}
	\quad ; \quad r = 0, \pm 1, \pm 2, \ldots \; . 
\label{eq:WSL}
\end{equation}
For instance, this can be the case when the force is mono\-chromatic with an 
additional static component. If
\begin{equation}
	F(t) = F_r + F_0\cos(\omega t) \; ,
\end{equation}
the static component~$F_r$ of the force ``tilts'' the lattice, leading to the 
appearance of Wannier-Stark ladders~\cite{Wannier60,AvronEtAl77}, and the 
equality~(\ref{eq:WSL}) gives
\begin{equation}
	F_r a = r \times \hbar\omega 
\end{equation} 
with $\omega = 2\pi/T$, requiring that the spacing $F_r a$ between the rungs 
of these ladders be equal to the energy of $r$~``photons''. While situations 
in which $r \ne 0$ do possess an intrinsic interest of their 
own~\cite{HolthausHone96,DreseHolthaus96}, here we restrict ourselves to 
$r = 0$.

The numerical solution of this eigenvalue equation~(\ref{eq:QBS}) can be 
accomplished by combining the techniques reviewed in Secs.~\ref{S_2} and 
\ref{S_3}. First one chooses a convenient and suitably truncated basis set 
which incorporates the spatial periodic boundary condition, and fixes a
wave number~$k$. Then each basis state is propagated in time over one 
period~$T$. The propagated states form the columns of the truncated one-cycle 
evolution matrix $U_k(T,0)$, the diagonalization of which finally yields both 
the approximate quasienergies $\varepsilon_n(k)$ ($\bmod \; \hbar\omega$), and 
the approximate expansion coefficients of the Floquet functions $u_{n,k}(x,0)$ 
with respect to the basis used. Repeating this procedure for a sufficiently
fine mesh of $k$-values gives the desired quasienergy dispersion relations.

\subsection{Ultracold atoms in shaken optical lattices}
\label{SubSec42}

In the particular case of ultracold atoms in an optical cosine lattice, as  
considered briefly in Sec.~\ref{S_2}, the time-periodic force on the 
electrically neutral particles is generated by exploiting their inertia. For
example, one can mount a mirror which retro-reflects a laser beam back 
into itself, and thus creates the standing wave transforming into the optical 
lattice potential, onto a piezoelectric actuator, enabling one to ``shake'' 
the lattice back and forth~\cite{ZenesiniEtAl09}. This shaking motion then 
transforms into an inertial force on the atoms in a frame comoving with the 
lattice. That is, one starts in the laboratory frame of reference with a 
Hamiltonian of the form  
\begin{equation}
	H^{\rm lab}(x,t) = \frac{p^2}{2M} 
	+ \frac{V_0}{2}\cos\big( 2\kL[x - \Delta L \cos(\omega t)]\big) \; ,
\label{eq:HSH}
\end{equation}
where $\Delta L$ denotes the amplitude of shaking, and performs the 
unitary transformation 
\begin{equation}
	\psi^{\rm lab}(x,t) = 
	\exp\left(-\frac{\ri}{\hbar}\Delta L \cos(\omega t) p \right)
	\psi^{\rm cm}(x,t) \; .
\end{equation}
Since, in analogy to Eq.~(\ref{eq:BCH}), one has the identity
\begin{eqnarray}
	& &
	\exp\left(+\frac{\ri}{\hbar} \Delta L \cos(\omega t) p \right) \, x \,
	\exp\left(-\frac{\ri}{\hbar} \Delta L \cos(\omega t) p \right)
\nonumber \\	& = & 
	x + \frac{\ri}{\hbar}\Delta L \cos(\omega t) [p,x]
\nonumber \\	& = &
	x + \Delta L \cos(\omega t) \; ,
	\phantom{\frac{\ri}{\hbar}}
\end{eqnarray}	 	
this brings us to the comoving frame of reference. Moreover, observing
that
\begin{eqnarray}
	& &
	\ri\hbar\frac{\rd}{\rd t} \psi^{\rm lab}(x,t) =  	
	\exp\left(-\frac{\ri}{\hbar} \Delta L \cos(\omega t) p \right)
\nonumber \\	& \times & 	
	\left( \ri\hbar\frac{\rd}{\rd t} \psi^{\rm cm}(x,t)
	 - \Delta L \omega \sin(\omega t) p \, \psi^{\rm cm}(x,t) \right) ,
\end{eqnarray}	 	
the Hamiltonian in this latter frame becomes		
\begin{eqnarray}
	H^{\rm cm}(x,t) & = &
	\frac{p^2}{2M} + \Delta L \omega \sin(\omega t) p
	+ \frac{V_0}{2} \cos(2\kL x)
\nonumber \\	& = &
	\frac{1}{2M}\Big( p + M \Delta L \omega \sin(\omega t) \Big)^2
	+ \frac{V_0}{2} \cos(2\kL x)
\nonumber \\	& &	
	- \frac{M}{2}(\Delta L \omega)^2 \sin^2(\omega t) \; ,	
\end{eqnarray}
having completed the square of the momentum in the second step. This 
comoving-frame Hamiltonian already resembles the previous 
operator~(\ref{eq:HXT}), except for its last, merely time-dependent
term. It might be tempting to ``gauge away'' this additional term by the 
further unitary transformation 
\begin{equation}
	\psi^{\rm cm}(x,t) = 
	\exp\left(\frac{\ri}{\hbar}\int_0^t \! \rd \tau \,
	\frac{M}{2}(\Delta L \omega)^2 \sin^2(\omega \tau) \right) \psi(x,t)
	\; ,
\label{eq:NPT}
\end{equation}
but this would be against the rules: Because of the form~(\ref{eq:FRU}) of
the time-evolution operator $U(t,0)$, a unitary transformation will leave the 
system's quasienergy spectrum unaltered only if the transformation itself is 
$T$-periodic. Now the elementary identity  
\begin{equation}
	\sin^2(\omega t) = \frac{1}{2} \Big( 1 - \cos(2\omega t) \Big) 
\end{equation}	
implies that the exponential generating the transformation~(\ref{eq:NPT})
contains a secular contribution which grows linearly in time, spoiling the
required $T$-periodicity. This deficiency is cured by extracting the secular 
term: Performing the $T$-periodic transformation 
\begin{eqnarray}
	\psi^{\rm cm}(x,t) & = &
	\exp\left(-\frac{\ri}{\hbar}\int_0^t \! \rd \tau \,
	\frac{M}{4}(\Delta L \omega)^2 \cos(2\omega \tau) \right) \psi(x,t)
\nonumber \\	& = &	
	\exp\left(-\frac{\ri}{8\hbar} M \Delta L^2 \omega
	\sin(2\omega t) \right) \psi(x,t) \; , 
\end{eqnarray}
we find that the wave functions~$\psi(x,t)$ are governed by the Hamiltonian
\begin{eqnarray}
	H(x,t) & = &
	\frac{1}{2M}\Big( p + M \Delta L \omega \sin(\omega t) \Big)^2
\nonumber \\	& &	
	+ \frac{V_0}{2} \cos(2\kL x)	
	- \frac{M}{4}(\Delta L \omega)^2 \; .	
\label{eq:HOL}
\end{eqnarray}
This corresponds to the previous form~(\ref{eq:HXT}), involving a sinusoidal 
force
\begin{equation}
	F(t) = F_0 \cos(\omega t)
\label{eq:SIF}
\end{equation}
with amplitude
\begin{equation}
	F_0 = M \Delta L \omega^2 \; .
\end{equation}
The additional energy shift appearing in the Hamiltonian~(\ref{eq:HOL}) 
possesses an intuitive interpretation: Consider a classical particle moving 
according to Newtons's equation
\begin{equation}
	M \ddot x(t) = F(t) \; ,
\end{equation}
so that
\begin{equation}
	\dot{x}(t) = \frac{F_0}{M\omega}\sin(\omega t) \; ,
\end{equation}
up to a constant. The average ``quiver energy'' of the particle then is
given by
\begin{eqnarray}
	\frac{1}{2} M \overline{{\dot{x}}^2} & = &
	\frac{F_0^2}{4M\omega^2}
\nonumber \\	& = &
	\frac{M}{4} (\Delta L \omega)^2 \; ;	
\label{eq:PON}
\end{eqnarray}				
this is the analog of the ``ponderomotive energy'' known from the study 
of electrons in intense laser fields~\cite{Gavrila92}. Thus, quasienergy 
spectra computed for a ``minimal coupling'' Hamiltonian~(\ref{eq:HXT}) 
incorporate the ponderomotive energy, while those computed for a ``shaking'' 
Hamiltonian~(\ref{eq:HSH}) do not.  

We now write the spatio-temporal Bloch waves~(\ref{eq:STB}) for a driven
optical lattice as
\begin{equation}
	\psi_{n,k}(x,t) = \re^{\ri k x}\chi_{n,k}(x,t) \; ,
\end{equation}
so that the Floquet states for given wave number~$k$, 
\begin{equation}
	\chi_{n,k}(x,t) = u_{n,k}(x,t) \re^{-\ri \varepsilon_n(k)t/\hbar} 
	\; ,
\end{equation}		
obey the Schr\"odinger equation 
\begin{widetext}
\begin{equation}
	\ri\hbar \frac{\rd}{\rd t} \chi_{n,k}(x,t) =
	\left[ \frac{1}{2M}
	\left(p + \hbar k + \frac{F_0}{\omega}\sin(\omega t) \right)^2
	+ \frac{V_0}{2}\cos(2 \kL x) \right] \chi_{n,k}(x,t) \; ,
\end{equation}
\end{widetext}
referring to the form~(\ref{eq:HXT}) of the Hamiltonian. We then employ 
the dimensionless scaled coordinate $z = \kL x$, as in Sec.~\ref{S_2},
together with the dimensionless time $\tau = \omega t$, and eliminate 
merely time-periodic terms without much ado. Dividing by the recoil 
energy~(\ref{eq:REC}), we arrive at the conveniently scaled equation
\begin{widetext}
\begin{equation}
	\ri \frac{\hbar\omega}{\ER} \frac{\rd}{\rd \tau} \chi_{n,k}(z,\tau)
	= \left[ -\frac{\rd^2}{\rd z^2} + \left( \frac{k}{\kL} \right)^2
	+ \frac{1}{2}\beta^2 + \frac{V_0}{2\ER} \cos(2z)
	+ 2 \left( \frac{k}{\kL} + \beta \sin \tau \right)
	\frac{1}{\ri} \frac{\rd}{\rd z}	\right] \chi_{n,k}(z,\tau)
\label{eq:NUM}
\end{equation}
\end{widetext}
which contains, apart from the reduced wave number $k/\kL$, three tunable
dimensionless parameters: The scaled lattice depth $V_0/\ER$, the scaled 
driving frequency $\hbar\omega/\ER$, and the scaled driving amplitude
\begin{equation}
	\beta = \frac{F_0/\kL}{\hbar\omega} \; .
\label{eq:BET}	
\end{equation}	
Observe that $\beta^2/2$ is the ponderomotive energy~(\ref{eq:PON}) in 
units of the recoil energy; this contribution should be left out when working 
with a ``shaking lattice'' Hamiltonian~(\ref{eq:HSH}). An efficient numerical 
algorithm for computing the quasienergy bands now again employs the 
trigonometric basis $\{ \varphi_\mu(z) \; ; \; \mu = 0,1,2,3,\ldots \}$ given 
by Eqs.~(\ref{eq:BA0}) and (\ref{eq:BAR}), so that the periodic boundary 
conditions in space are properly accounted for, and the operators entering 
into this Eq.~(\ref{eq:NUM}) aquire the matrix representations~(\ref{eq:MX1}),
(\ref{eq:MX2}), and (\ref{eq:MX3}), suitably truncated. The actual computation 
then proceeds in precise analogy to the example of the driven particle in the 
box studied in Chap.~\ref{SubSec33}: Each basis state $\varphi_\mu(z)$ is taken as 
initial condition $\chi(z,0)$ for Eq.~(\ref{eq:NUM}) and propagated over one 
period in time, providing the $\mu$th column of the mono\-dromy matrix; upon 
diagonalization, this matrix yields quasienergies and Floquet functions. As 
shown in the Appendix, the numerical effort can still be reduced by exploiting 
the symmetries of the quasienergy operator.

\subsection{Multiphoton-like resonances in optical lattices}
\label{SubSec43}

To start the discussion of actual Floquet engineering, and to give some 
selected examples of the rich variety of quasienergy dispersion relations 
that can be realized with driven optical lattices, we again consider cosine 
potentials with depth $V_0/\ER = 4.0$ or $V_0/\ER = 8.0$, for which the 
undriven energy dispersion relations had been displayed in Fig.~\ref{F_2}; 
some relevant figures of merit for these lattices are listed in Table~\ref{T_1}. 
For orientation: If one works with atoms of $^{87}$Rb in a lattice generated 
by laser radiation with wave length $\lambda = 842$~nm, as in an experiment 
reported by Zenesini {\em et al.\/}~\cite{ZenesiniEtAl09}, the recoil frequency
$\nu = \ER/h$ amounts to 3.23~kHz; typical driving frequencies thus fall into 
the lower kilohertz regime.

\begin{table}[b]
\begin{tabular}{||c|c|c|c||} \hline
$~V_0/\ER~$  &  $~W_0/\ER~$  &  $~\Delta_{01}/\ER~$  &  $~W_1/\ER~$  \\ \hline
    4.0      &      0.345    &              1.969    &      2.058    \\ 	
    8.0      &      0.123    &		    3.770    &      1.293    \\ \hline
\end{tabular}
\caption{Widths $W_0$ and $W_1$ of the lowest two energy bands, and
	magnitude of the gap $\Delta_{01}$ between them at $k/\kL = \pm 1$, 
	for optical lattices with the depths $V_0$ considered in 
	Fig.~\ref{F_2}.} 
\label{T_1}
\end{table}

\begin{figure}
\includegraphics[width = 0.9\linewidth]{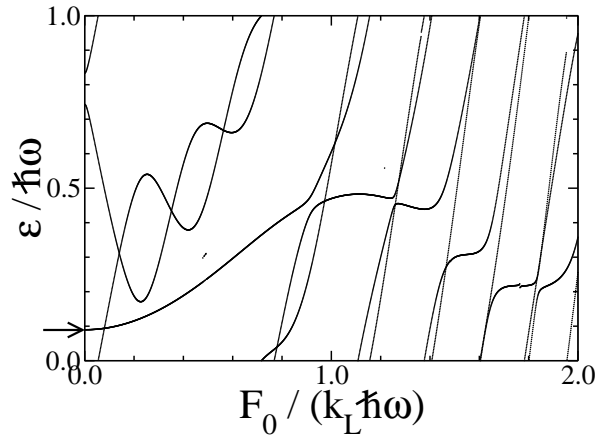}
\caption{ac-Stark shifts of the edges $k/\kL = 0$ of the lowest three bands 
	$n = 0$ (indicated by the arrow), $n = 1$, and $n=2$ of an optical 
	cosine lattice with depth $V_0/\ER = 4.0$ in response to a driving 
	force with frequency $\hbar\omega/\ER = 0.5$.}  
\label{F_6}	
\end{figure}

\begin{figure}
\includegraphics[width = 0.9\linewidth]{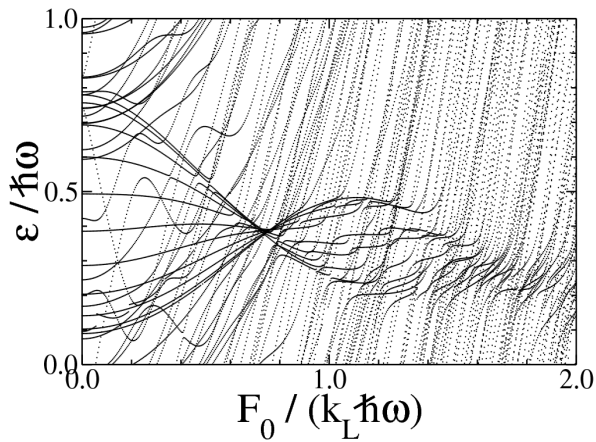}
\caption{Quasienergy band originating from the lowest energy band for 
	$V_0/\ER = 4.0$ and $\hbar\omega/\ER = 0.5$, embedded in a host of 
	quasienergies from other bands. Shown are quasienergy eigenvalues 
	with reduced wave numbers~$k/\kL$ varying between $0.0$ and $1.0$ in 
	steps of $0.1$. To produce this figure, the Floquet states have been
	ordered according to their overlap with the basis states, and only
	quasienergies belonging to the respective three ``lowest'' Floquet 
	states have been plotted. In the coarse-graining regime where the 
	band is still recognizable as such, Floquet wave packets~(\ref{eq:WPA})
	are protected against heating, whereas the system would heat up soon 
	in the resonance-dominated regime.}  
\label{F_7}	
\end{figure}

In Fig.~\ref{F_6} we show the ac-Stark shifts of the edges $k/\kL = 0$ of the 
lowest three bands of a cosine lattice with depth~$V_0/\ER = 4.0$ in response 
to a drive applied with the frequency $\hbar\omega/\ER = 0.5$, so that the 
``photon'' energy $\hbar\omega$ is somewhat larger than the width~$W_0$ of the 
undriven lowest energy band, while the gap~$\Delta_{01}$ between the lowest 
and the first excited band is roughly equal to $4 \, \hbar\omega$. Despite 
this gap, the shifting ``lowest'' band edge, indicated by the arrow, is heavily 
disrupted by resonances with the shifted higher edges when the driving 
amplitude~(\ref{eq:BET}) reaches the nonperturbative regime. Note that this 
figure includes the ponderomotive energy, as do all following ones. But still, 
in order to assess the dynamics of a Floquet wave packet~(\ref{eq:WPA}) the 
inspection of states with just one single wave number is not sufficient; one 
rather has to find a way of visualizing the response of entire quasienergy 
bands to the drive. To this end, we have combined in Fig.~\ref{F_7} 
quasienergies with 11 reduced wave numbers $k/\kL = 0.0$, $0.1$, \ldots, $1.0$,
for the same parameters as employed in Fig.~\ref{F_6}. After each solution 
of the eigenvalue equation for the respective prescribed values of $k/\kL$ 
and $\beta$ the Floquet states have been ordered according to their overlap 
with the basis states as discussed for the particle in the box in 
Chap.~\ref{SubSec33}, and only the quasienenergies of the respective three 
``lowest'' Floquet states have been included in the plot. Because of the 
multitude of avoided crossings involved, this procedure cannot always give 
smooth lines, but it clearly brings out the key features: The {\em internal\/} 
ac-Stark shift, that is, the ac-Stark shift of states with different $k/\kL$ 
within the same band relative to each other, effectuates a narrowing of the 
quasienergy band emerging from the lowest energy band $n = 0$ when $\beta$ 
is increased from zero, such that the width of this quasienergy band approaches
zero when $\beta \approx 0.74$. The width increases again when $\beta$ is 
enlarged still further, but for $\beta > 1.0$ the band is virtually torn apart 
by a plethora of multiphoton resonances with ac-Stark-shifted other bands. 
This Fig.~\ref{F_7} also exemplifies what the concept of coarse graining, as 
introduced in Chap.~\ref{SubSec35}, means in practice: In the coarse-graining 
regime, where the quasienergy band remains almost intact, a Floquet wave 
packet~(\ref{eq:WPA}) prepared in this band hardly couples to higher ones, so 
that the system is well protected against heating, whereas it would be subject 
to rapid heating in the resonance-dominated regime.

Since the gap $\Delta_{01}$ here amounts to the energy of about four 
``photons'', the driven lowest energy band remains reasonably well isolated 
from the higher ones at least for low driving amplitudes. In this case one can 
resort to a single-band approximation: Provided the lattice is so deep that 
the lowest Wannier state localized in one of the lattice wells has appreciable 
overlap with its nearest neighbors only~\cite{Kohn59}, the unperturbed lowest 
energy band acquires a cosine form, 
\begin{equation}
	E_0(k) = E_c - \frac{W_0}{2}\cos(ka) \; ,          
\label{eq:ICB}
\end{equation}
where~$a$ is the lattice constant. In the case of optical cosine lattices 
the relative error of this approximation still is on the order of 7\% when 
$V_0/\ER = 4.0$, and drops to 2\% when $V_0/\ER = 8.0$~\cite{EckardtEtAl09,
BoersEtAl07}. When an isolated cosine band~(\ref{eq:ICB}) is driven by a 
sinusoidal force~(\ref{eq:SIF}) the calculation of the resulting quasienergy 
band becomes elementary, giving~\cite{EckardtEtAl09,Holthaus92,Holthaus92b}  
\begin{equation}
	\varepsilon_0(k) = E_c - \frac{W_0}{2} 
	J_0\left(\frac{F_0 a}{\hbar\omega}\right)\cos(ka)  
   	\quad \bmod \hbar\omega \; .
\label{eq:IQB}
\end{equation}	   
Here $J_0(z)$ is the Bessel function of order zero. Starting from $J_0(0) = 1$ 
this function oscillates with increasing real argument~$z$, with an 
amplitude decreasing as $1/\sqrt{z}$; its first two zeros are located at 
$z_1 \approx 2.405$ and $z_2 \approx 5.520$~\cite{AbramowitzStegun70b}.
Thus, the result~(\ref{eq:IQB}) derived for a perfectly isolated band explains 
part of the observations made in Fig.~\ref{F_7}. At the zeros of the Bessel 
function the width of the band~(\ref{eq:IQB}) vanishes, so that it becomes 
dispersionless. Since $a = \pi/\kL$ for optical cosine lattices, we have 
\begin{equation}
	\frac{F_0 a}{\hbar\omega} = \pi\beta \; ,
\end{equation}
implying that the first band flattening can occur when the scaled driving 
amplitude~(\ref{eq:BET}) adopts the value
\begin{equation}
	\beta_1 = z_1/\pi \approx 0.765 \; ; 
\label{eq:CP1}
\end{equation}
the second ``collapse point'' is located at
\begin{equation}
	\beta_2 = z_2/\pi \approx 1.757  
\label{eq:CP2}
\end{equation}
in the ideal case.

\begin{figure}
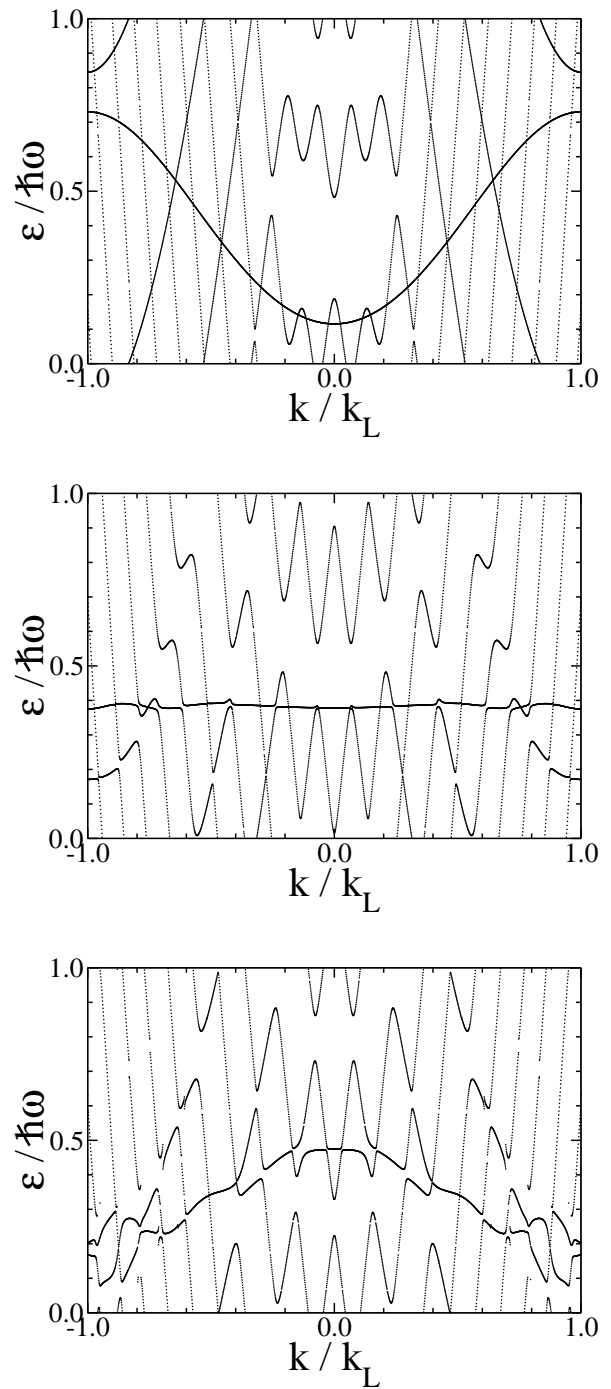

\includegraphics[width = 0.9\linewidth]{QES_Fig8a.eps}

\vspace*{4ex}

\includegraphics[width = 0.9\linewidth]{QES_Fig8b.eps}

\vspace*{4ex}

\includegraphics[width = 0.9\linewidth]{QES_Fig8c.eps}
\caption{Quasienergy dispersion relations for $V_0/\ER = 4.0$ and 
	 $\hbar\omega/\ER = 0.5$, implying that the gap $\Delta_{01}$ between 
	 the lowest and the first excited unperturbed energy band is a bit
	 smaller than $4 \, \hbar\omega$ (see Table~\ref{T_1}).   	 
	 For scaled driving amplitude $\beta = 0.20$ the quasienergy band
	 originating from the lowest energy band still is well described by 
	 the single-band approximation~(\ref{eq:IQB}) (upper panel). For
	 $\beta = 0.74$, corresponding to the ``collapse point'' observed
	 in Fig.~\ref{F_7}, the band is not perfectly flat, but rather
	 disrupted by several small anticrossings (middle panel). For
	 $\beta = 1.21$, for which an isolated band~(\ref{eq:IQB}) would
	 be maximally inverted, strong hybridization occurs (lower panel),
	 forecasting rapid heating.} 
\label{F_8}	
\end{figure}

In contrast to the assumptions underlying the single-band 
expression~(\ref{eq:IQB}), a driven energy band in an optical lattice is not 
isolated, but subject to interband transitions. Hence, the quasienergy band 
emerging from the lowest energy band incorporates these transitions by acquiring
admixtures from higher energy bands. Even if the energy gap $\Delta_{01}$ 
amounts to the energy of several ``photons'', so that, perturbatively speaking,
it can only be bridged by higher-order multiphoton transitions, this 
hybridization is non-negligible when the scaled driving amplitude~$\beta$ 
becomes nonperturbatively strong. This is precisely what is seen in 
Fig.~\ref{F_7}: For $\beta < 1$ multiphoton resonances still play a minor role,
so that Eq.~(\ref{eq:IQB}) describes the exact numerical data reasonably well 
in this interval, in the ``coarse graining'' sense put forward in 
Chap.~\ref{SubSec35}. In particular, an approximate band collapse is observed 
at $\beta \approx 0.74$, quite close to the predicted value~(\ref{eq:CP1}). 
But for $\beta > 1.5$ the resonances become so strong that the band is almost
completely disrupted, so that coarse graining becomes impossible and a second 
band collapse cannot be realized. The increasing importance of multi\-photon 
resonances with increasing~$\beta$ becomes particularly evident in 
Fig.~\ref{F_8}, where we have visualized the quasienergy dispersion relation 
of the quasienergy band emerging from the ground-state energy band; again 
these plots have been produced by selecting the ``lowest'' three quasienergies 
for each value of the reduced wave number~$k/\kL$. In the upper panel of 
Fig.~\ref{F_8} we have $\beta = 0.2$; here the cosine band~(\ref{eq:IQB}) is 
well preserved. It should be kept in mind, however, that the coarse graining 
philo\-sophy is implied already here: Lots of (in fact, infinitely many)
non-resolved anticrossings are tacitly swept under the carpet. In the middle 
panel we have $\beta = 0.74$, corresponding to the smallest band width found 
in Fig.~\ref{F_7}. Evidently the quasienergy band is ``essentially'' flat, 
but it is already visibly affected by quite a number of narrow resonances. 
When $\beta = 1.21$ the ideal band~(\ref{eq:IQB}) would be maximally inverted, 
because the Bessel function takes on its largest negative value here. But, as 
seen in the lower panel of Fig.~\ref{F_8}, the actual quasienergy band then 
hybridizes strongly, developing wide anticrossings which render a single-band 
description useless. In experimental terms, the widths of these anticrossings 
quantify the strength of the coupling of the system's wave function to higher 
states, assuming that it had initially been prepared in the lowest band, 
and thus determine the time scales associated with heating: While a weakly 
interacting Bose gas in a driven optical lattice remains shielded against 
heating in the coarse-graining regime for hundreds or thousands of driving 
cycles~\cite{LignierEtAl07,EckardtEtAl09}, it would heat up much faster, and 
be lost from the lattice, when the resonances start to make themselves felt.

Hence, one of the appealing prospects offered by driven optical lattices is 
that one can easily reach the realm of nonperturbatively strong driving, 
which would not be accessible in an equally clean manner with laser-driven 
electrons in crystal lattice potentials, say. The scaled driving 
amplitude~(\ref{eq:BET}) is related to the shaking amplitude~$\Delta L$ 
entering into the laboratory-frame Hamiltonian~(\ref{eq:HSH}) through   
\begin {equation}
	\beta = \frac{\pi}{2} \, \frac{\hbar\omega}{\ER} \, 
	\frac{\Delta L}{a} \; ,  
\label{eq:BDL}
\end{equation}	
so that values of $\beta$ on the order of unity can be reached when~$\Delta L$ 
is comparable to the lattice constant~$a$. In this sense, driven optical 
lattices filled with ultracold atoms may serve as ``strong field simulators'', 
allowing one to probe multiphoton transitions in a periodic potential without 
encountering the many competing charge-induced effects that would mask them 
in solid-state samples interacting with laser fields of equivalent
strengths~\cite{ArlinghausHolthaus10,ArlinghausHolthaus12}.

\subsection{Isolating a single cosine quasienergy band}
\label{SubSec44}

\begin{figure}[t]
\includegraphics[width = 0.9\linewidth]{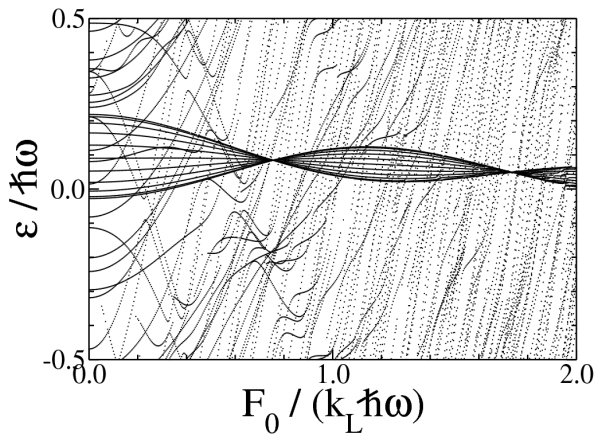}
\caption{Quasienergy band for $V_0/\ER = 8.0$ and $\hbar\omega/\ER = 0.5$,
	so that $\Delta_{01} = 7.54 \, \hbar\omega$ here. This plot has 
	been produced in the same manner as Fig.~\ref{F_7}.}
\label{F_9}	
\end{figure}

\begin{figure}[t!]
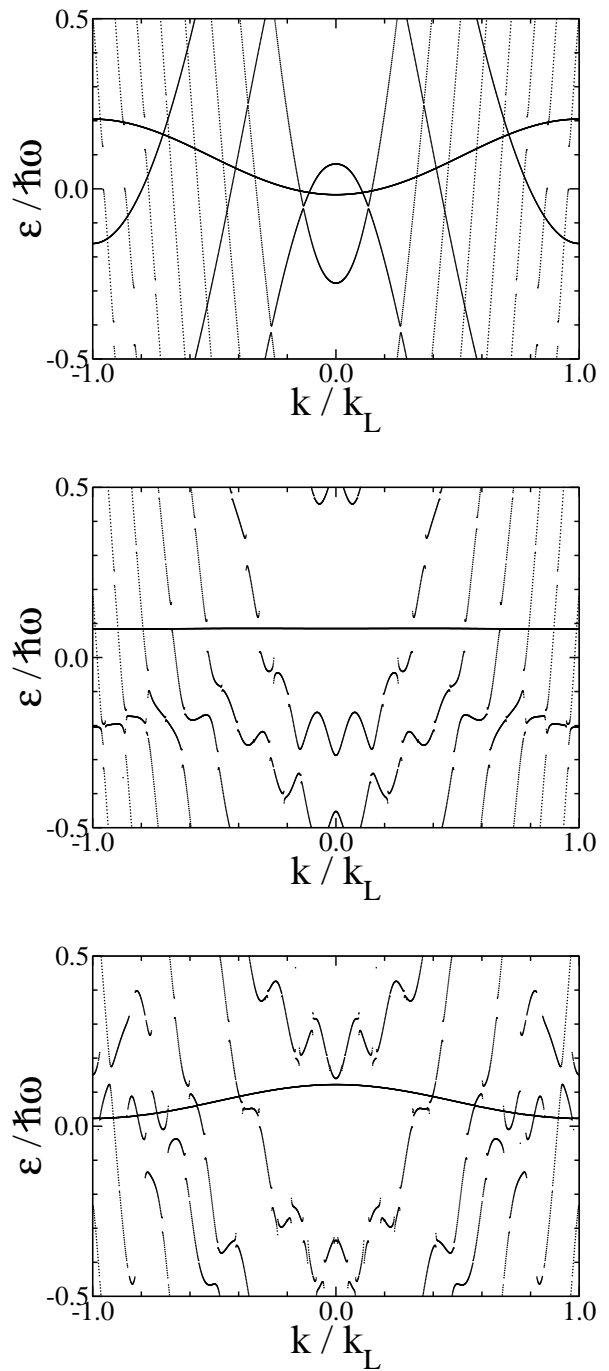

\includegraphics[width = 0.9\linewidth]{QES_Fig10a.eps}

\vspace*{4ex}

\includegraphics[width = 0.9\linewidth]{QES_Fig10b.eps}

\vspace*{4ex}

\includegraphics[width = 0.9\linewidth]{QES_Fig10c.eps}
\caption{Quasienergy dispersion relations for $V_0/\ER = 8.0$ and
	$\hbar\omega/\ER = 0.5$. For $\beta = 0.2$ (upper panel),
	$\beta =0.76$ (middle panel), and $\beta = 1.21$ (lower panel) 
	the coarse-grained quasienergy band emerging from the lowest energy 
	band is well described by the single-band formula~(\ref{eq:IQB}).} 	 
\label{F_10}	
\end{figure}

If, however, one desires to come closer to the single-band ideal~(\ref{eq:IQB})
than in the previous example, and to engineer practically flat or perfectly
inverted cosine bands, the ``degree of isolation'' of the ground-state energy
band has to be enhanced. This can be achieved, for instance, by increasing the 
lattice depth, which amounts to an enlargement of the gap $\Delta_{01}$, while
keeping the driving frequency at its previous value. Following that strategy,
Fig.~\ref{F_9} depicts the ``lowest'' quasienergy band for $V_0/\ER = 8.0$
and $\hbar\omega/\ER = 0.5$. According to Table~\ref{T_1} this implies
$\Delta_{01}/(\hbar\omega) = 7.54$, so that now more than seven ``photons'' are
required to bridge the gap, whereas it had been about four in Fig.~\ref{F_7}.
Consequently one now obtains a quasienergy band which remains more or less
intact even up to $\beta = 2.0$, with two collapse points at the predicted 
positions~(\ref{eq:CP1}) and~(\ref{eq:CP2}). The validity of Eq.~(\ref{eq:IQB})
is confirmed in greater detail by Fig.~\ref{F_10}: The cosine band starts to
narrow when $\beta = 0.2$ (upper panel), becomes perfectly flat --- at least on
the scale of Fig.~\ref{F_10}; coarse graining still is implied here! --- when
$\beta = 0.76$ (middle panel), and is maximally inverted when $\beta = 1.21$
(lower panel). It is interesting to observe how the attempted flattening of a 
second band at $\beta = 0.76$ is thwarted by an aligned sequence of broad 
resonances.

The feasibility of close-to-perfect band inversion in real optical 
lattices has a noteworthy consequence: A weakly interacting Bose-Einstein 
condensate occupies the state with $k/\kL = \pm 1$ when the band is 
inverted~\cite{ArimondoEtAl12}, instead of the usual ground state $k/\kL = 0$ 
of an undriven cosine band~(\ref{eq:ICB}). It also deserves to be mentioned 
that first indications for band narrowing in periodically driven optical 
lattices loaded with cold, but non-condensed atoms had been observed in 
pioneering experiments as early as 1998~\cite{MadisonEtAl98}, but the Bessel 
function could not be mapped out cleanly beyond the first collapse point 
until a decade later through experiments with driven Bose-Einstein 
condensates~\cite{LignierEtAl07,EckardtEtAl09}.

Thus, for carefully selected parameters the quasienergy band which emerges 
in a sinusoidally driven optical lattice from its lowest energy band is well 
described by Eq.~(\ref{eq:IQB}). This observation creates a link to the subject
of dynamic localization: It had been theoretically discussed already in 1986 
that a charged particle moving under the influence of an oscillating electric 
field on a defect-free single-band tight-binding lattice with equal couplings 
between neighboring sites remains ``dynamically'' localized when the parameter 
$F_0 a/(\hbar\omega)$ equals a zero of the Bessel function $J_0$, with $F_0$ 
denoting the product of the particle's charge and the amplitude of the 
electric field~\cite{DunlapKenkre86}. The introduction of quasienergy 
bands~\cite{Holthaus92,Holthaus92b} allows one to understand this remarkable 
phenomenon within the established concepts of solid-state physics: If one 
builds an arbitrary  wave packet~(\ref{eq:WPA}) from the spatio-temporal Bloch 
waves which form a quasienergy band~(\ref{eq:IQB}) and follows its evolution 
in time, the componets of this packet generally will dephase, so that the 
packet generally will delocalize, except for those para\-meters for which the 
band is dispersionless, {\em i.e.\/}, when $F_0 a/(\hbar \omega)$ equals a 
zero of $J_0$. Then all Floquet components of the wave packet acquire precisely
the same phase factor after each driving cycle, so that the packet reproduces 
itself perpetually. Hence, dynamic localization results from prohibited 
dephasing~\cite{ArlinghausEtAl11}. 

Since the width of an ideal cosine band is determined by the hopping matrix 
elements between neighboring sites, the appearance of the Bessel function~$J_0$
in the quasienergy band~(\ref{eq:IQB}) may be interpreted as an effective 
renormalization of these hopping matrix elements caused by the oscillating 
force. This view leads to a number of further applications. For instance, 
a tight-binding chain with quasiperiodically varying on-site energies exhibits 
a metal-insulator transition at a certain critical ratio of the quasiperiodic 
perturbation strength and the hopping strength~\cite{AubryAndre80,Sokoloff85}. 
Since the driving-induced renormalization of the hopping strength allows one
to tune this ratio by applying an oscillating force, one encounters a new
mechanism for switching from one phase to the other: The magnitude of the 
driving amplitude decides whether the system is in the ``metallic'' or in the 
``insulating'' phase; this engineering option can be explored with ultracold 
atoms in driven bichromatic optical lattices~\cite{DreseHolthaus97b,
ArlinghausEtAl11,DreseHolthaus97a}. By the same token, the average extension 
of Anderson-localized states in a randomly perturbed lattice is determined by 
the ratio of the typical strength of the random on-site perturbation and the 
hopping strength; accordingly, the degree of Anderson localization can be 
controlled by an external oscillating force~\cite{HolthausEtAl95}. Of 
particular significance is the recognition that the concept of hopping-strength 
renormalization survives the presence of interparticle interaction in many-body
systems~\cite{EckardtEtAl05,EckardtHolthaus07}. Namely, the prototypal quantum 
phase transition from a superfluid to a Mott insulator undergone by a gas of 
ultracold atoms in an optical lattice~\cite{GreinerEtAl02} is governed by the 
ratio of the on-site interaction strength and the nearest-neighbor hopping 
strength. Therefore, the phase boundary can be crossed in a periodically 
driven optical lattice by varying the driving amplitude~\cite{EckardtEtAl05},
as has been confirmed precisely in a path-directing experiment by Zenesini 
{\em et al.\/}~\cite{ZenesiniEtAl09}. In line with this proof of principle, 
related theoretical proposals for controlling the phase transition have been 
made~\cite{EckardtHolthaus07,CreffieldMonteiro06}, and the driving-induced 
renormalization of the hopping strength has been exploited still further 
in a recent series of quite different innovative optical-lattice 
experiments~\cite{StruckEtAl11,StruckEtAl12,StruckEtAl13,AidelsburgerEtAl13,
MiyakeEtAl13,JotzuEtAl14}. A concise discussion of the current status of these 
topical developments is given by Eckardt~\cite{Eckardt15}.

\subsection{Floquet engineerig with interband ac-Stark shifts}
\label{SubSec45}

\begin{figure}[th!]
\includegraphics[width = 0.9\linewidth]{QES_Fig11a.eps}

\vspace*{4ex}

\includegraphics[width = 0.9\linewidth]{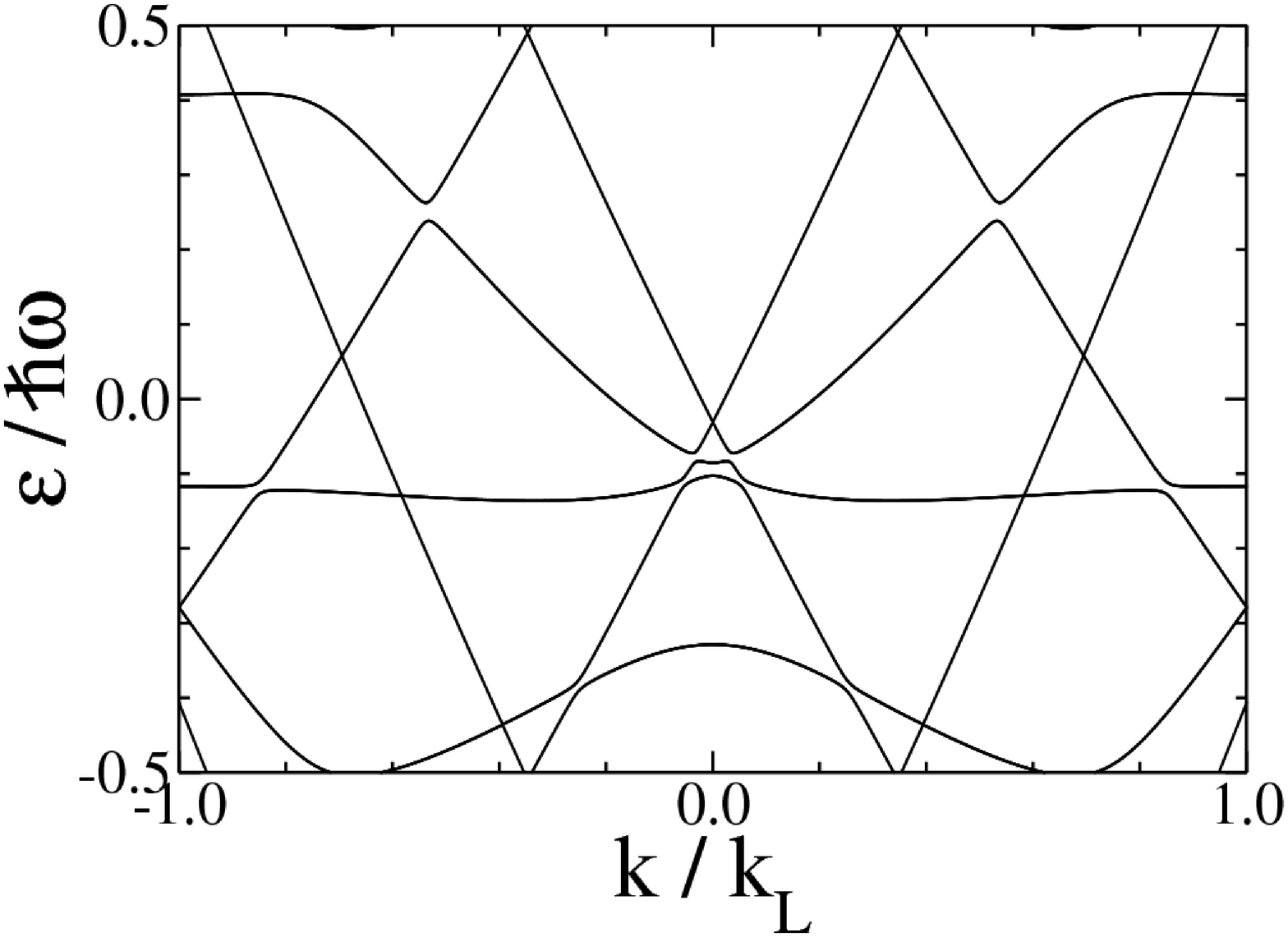}

\vspace*{4ex}

\includegraphics[width = 0.9\linewidth]{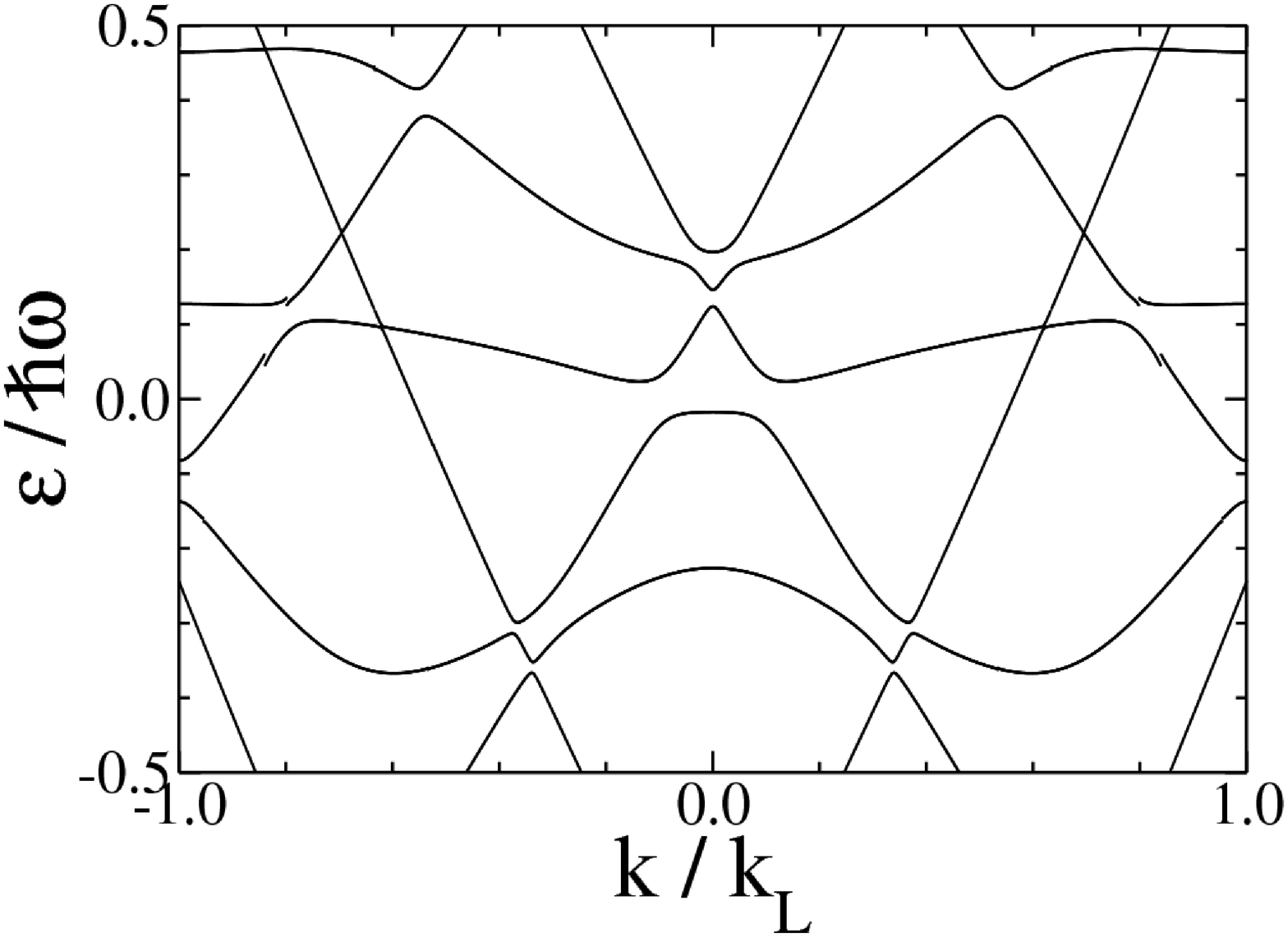}
\caption{Quasienergy dispersion relations for $V_0/\ER = 7.0$ and 
	$\hbar\omega/\ER = 5.51$, as corresponding to Ref.~\cite{ParkerEtAl13}. 
	For $\beta = 0.17$ (upper panel) the ``lowest'' quasienergy band 
	$n = 0$, pinched between the bands $n = 1$ (below) and $n = 2$ (above) 
	in the quasienergy Brilloiun zone, exhibits a shallow double-well 
	structure, as magnified in Fig.~\ref{F_12}. For $\beta =0.69$ (middle 
	panel) the ac-Stark-shifted ground-state band resonates with driven 
	free particle-like above-barrier states. For $\beta = 1.50$ (lower 
	panel) the avoided crossings have become wider, leading to an even 
	more pronounced double-well structure close to the center of the 
	quasimomentum Brillouin zone.}	 
\label{F_11}	
\end{figure}

Flat-band engineering and band inversion, as exemplified in Fig.~\ref{F_10},
rely on the internal ac-Stark effect within a single band. The exploitation
of the {\em interband\/} ac-Stark effect, that is, of the driving-induced 
shift of different bands relative to each other, opens up further options. 
A seminal example of this type of Floquet engineering has been provided by 
Parker {\em et al.\/}~\cite{ParkerEtAl13}: After loading a Bose-Einstein 
condensate of $^{133}$Cs atoms into an optical lattice created with laser 
wavelength $\lambda = 1064$~nm, amounting to the recoil frequency
$\ER/h = 1.325$~kHz, these authors have coupled the lowest two energy bands 
by employing a driving frequency which is blue-detuned from a transition 
between them: Working with a lattice of depth $V_0/\ER = 7.0$, one finds 
the scaled band separation $E_1(0) - E_0(0)= 4.96 \, \ER$ at $k/\kL = 0$, 
while $E_1(\pm 1) - E_0(\pm 1) = 3.34 \, \ER$ at $k/\kL = \pm 1$; 
shaking that lattice with frequency $\omega/(2\pi) = 7.3$~kHz, or 
$\hbar\omega/\ER = 5.51$, then gives a blue-detuning by about 11\% at
$k/\kL = 0$, but by about $65\%$ at $k/\kL = \pm 1$~\cite{ParkerEtAl13}.

\begin{figure}[t]
\includegraphics[width = 0.9\linewidth]{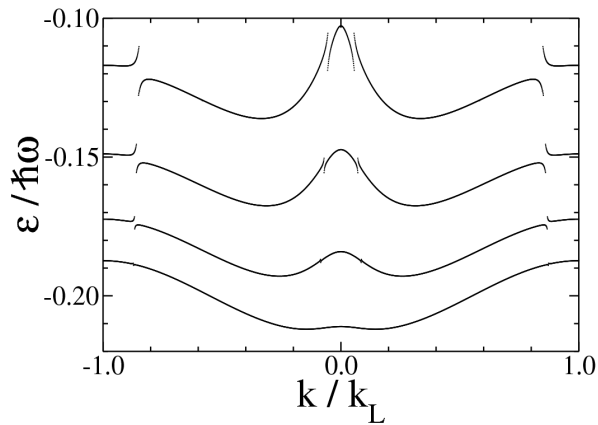}
\caption{ac-Stark-shifted quasienergy band emerging from the ground-state
	energy band of an optical lattice with depth $V_0/\ER = 7.0$ under
	driving with frequency $\hbar\omega/\ER = 5.51$. The driving
	amplitudes~$\beta$ are $0.17$, $0.35$, $0.52$, and $0.69$ (bottom 
	to top). The avoided crossings result from multi\-photon resonances 
	with the band $n = 3$, providing a potential escape channel. This 
	figure corresponds precisely to Fig.~S2 shown in the Supplementary 
	Information of Ref.~\cite{ParkerEtAl13}.} 
\label{F_12}	
\end{figure}

As shown in the upper panel of Fig.~\ref{F_11}, for low driving amplitudes
this choice of parameters places the lowest band $n = 0$ half-way between 
the first excited band $n = 1$ and the second excited band $n = 2$ in the 
quasienergy Brillouin zone. For moderate driving strength one may then regard 
the coupled lowest two bands $n = 0$ and $n = 1$ as set of two-level systems, 
parametrized by the wave number $k/\kL$; each one behaving in accordance with 
the RWA-formula~(\ref{eq:QTM}). Recalling the assignment~(\ref{eq:LI2}) for 
blue detuning, the quasienergy representatives connected to the ground-state 
band $n = 0$ then are shifted upwards with increasing driving amplitude, 
whereas the representatives connected to the excited band $n = 1$ are shifted 
downwards. Now the detuning varies throughout the quasimomentum Brillouin zone,
being smallest at its center $k/\kL = 0$ and being largest at the edges 
$k/\kL = \pm 1$. According to the expression~(\ref{eq:GRF}) for the generalized
Rabi frequency, this implies that the upwards-ac-Stark shift of the lowest band 
is largest at the zone center, and smallest at the edges; in combination with 
the already existing cuvature of the unperturbed energy band these unequal 
shifts result in the formation of a double well-shaped quasienergy dispersion 
relation. This mechanism is illustrated in Fig.~\ref{F_12} for the parameters 
employed in the experiment~\cite{ParkerEtAl13}; with the help of the 
relation~(\ref{eq:BDL}) the scaled driving strengths $\beta = 0.17$, $0.35$, 
$0.52$, and $0.69$ are converted into the peak-to-peak shaking amplitudes 
$\Delta x = 2\,\Delta L = 21$~nm, 43~nm, 64~nm, and 85~nm, respectively.

\begin{figure}[t]
\includegraphics[width = 0.9\linewidth]{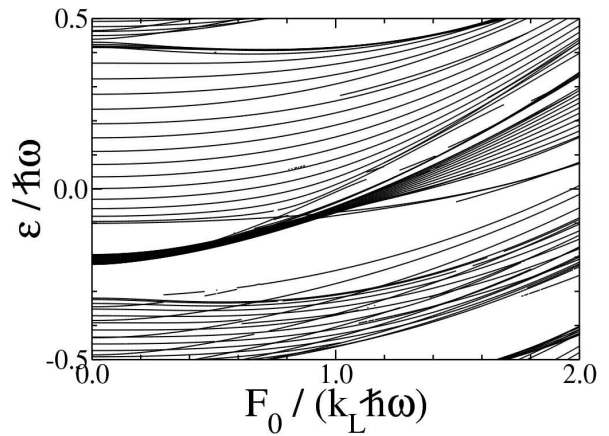}
\caption{Lowest three quasienergy bands for $V_0/\ER = 7.0$ and 
	$\hbar\omega/\ER = 5.51$ (cf.\ Fig.~\ref{F_11}); each band is 
	represented by 21 scaled wave 
	numbers~$k/\kL$ ranging from $0$ to $1$ in equal steps of $0.05$. 
	Observe how the quasienergy band $n = 0$ widens after touching its 
	upper partner $n = 2$, whereas that latter band becomes more narrow, 
	indicating stong ``vertical'' hybidization of both bands.}
\label{F_13}	
\end{figure}

This double well-shaped quasienergy dispersion is of profound experimental
significance, since a driven Bose gas condenses into one of the two wells,
allowing one to simulate ferromagnetism~\cite{ParkerEtAl13}; if 
interparticle interactions are taken into account by means of a Bogoliubov
treatment, one obtains a ``roton-maxon'' dispersion as known from 
He~II~\cite{HaEtAl15}. So far one works with relatively low shaking 
amplitudes, avoiding the resonances which show up strongly in Fig.~\ref{F_12} 
in the center of the barrier for $\beta = 0.69$; the expanded view provided
by the middle panel of Fig.~\ref{F_11} allows one to trace the anticrossing
partner to the ``above barrier''-band $n = 3$. However, for substantially 
higher driving amplitudes such as $\beta = 1.5$ considered in the lower panel 
of Fig.~\ref{F_11}, the quasienergy bands have disentangled themselves, 
creating a quite pronouned, smooth  double-well structure in the center of 
the quasimomentum Brillouin zone. Thus, if one succeeds in populating the 
corresponding quasienergy band, the strongly driven system may actually be more 
stable than one with lower driving amplitudes, but one has to find a way to 
protect the atoms from the intermediate resonances encountered when the drive 
is turned on.   

Another peculiar strong-driving phenomenon is depicted in Fig.~\ref{F_13}:
Similar to Figs.~\ref{F_7} and~\ref{F_9}, here we have plotted the 
quasienergies emanating from the three lowest energy bands vs.\ the scaled
driving amplitude, now for 21~wave numbers $k/\kL$ ranging from~$0$ to~$1$
in steps of $0.05$. Interestingly, the bands $n = 0$ and $n = 2$ touch
at $\beta \approx 0.8$, and then hybridize strongly: The quasienergy band
$n = 0$ fans out and widens, whereas its partner $n = 2$ becomes more narrow.
This type of resonant band hybridization is the subject of the following
chapter.

\subsection{Floquet engineering with resonances}
\label{SubSec46}

\begin{figure}[th!]
\includegraphics[width = 0.9\linewidth]{QES_Fig14a.eps}

\vspace*{4ex}

\includegraphics[width = 0.9\linewidth]{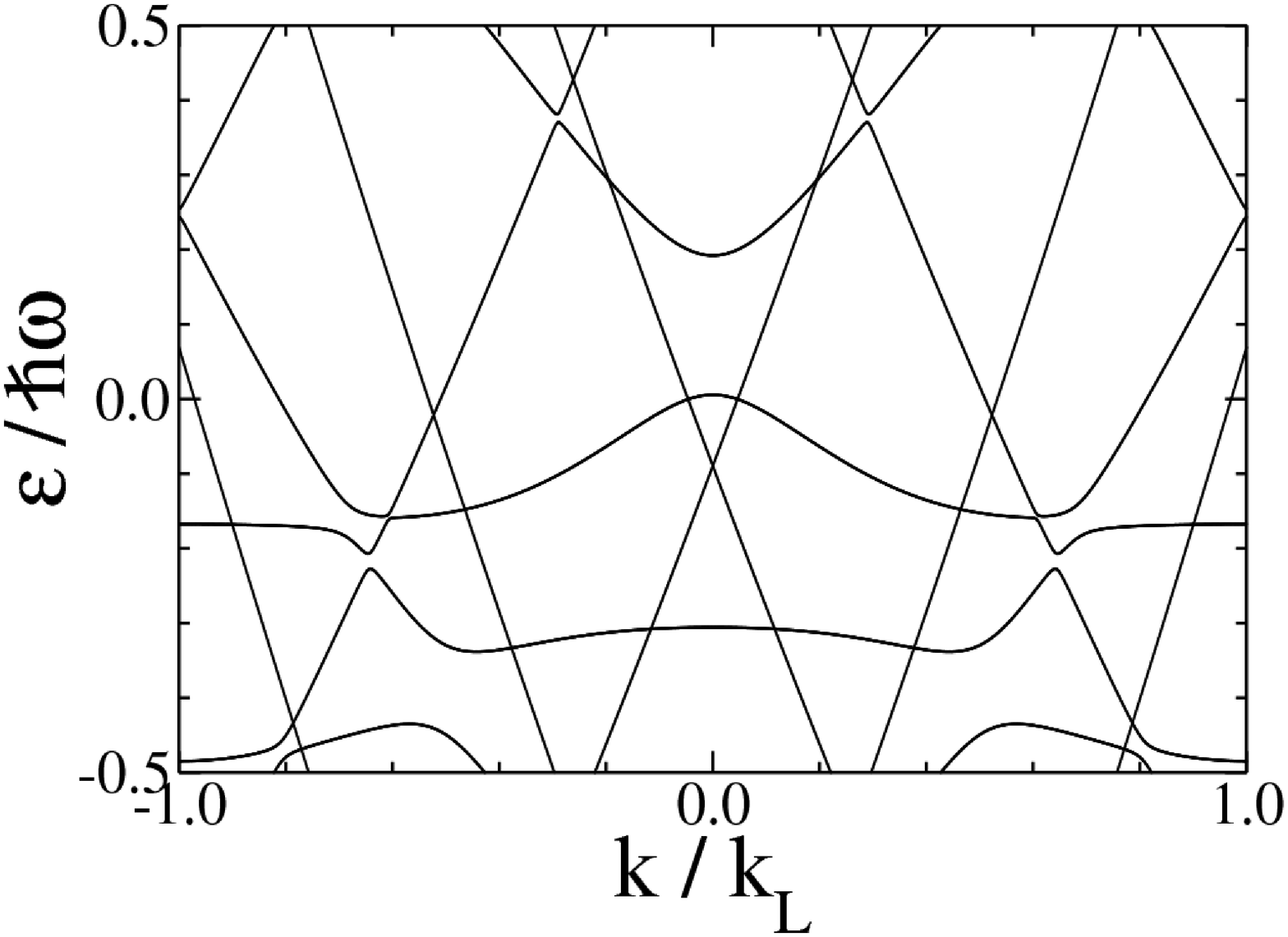}

\vspace*{4ex}

\includegraphics[width = 0.9\linewidth]{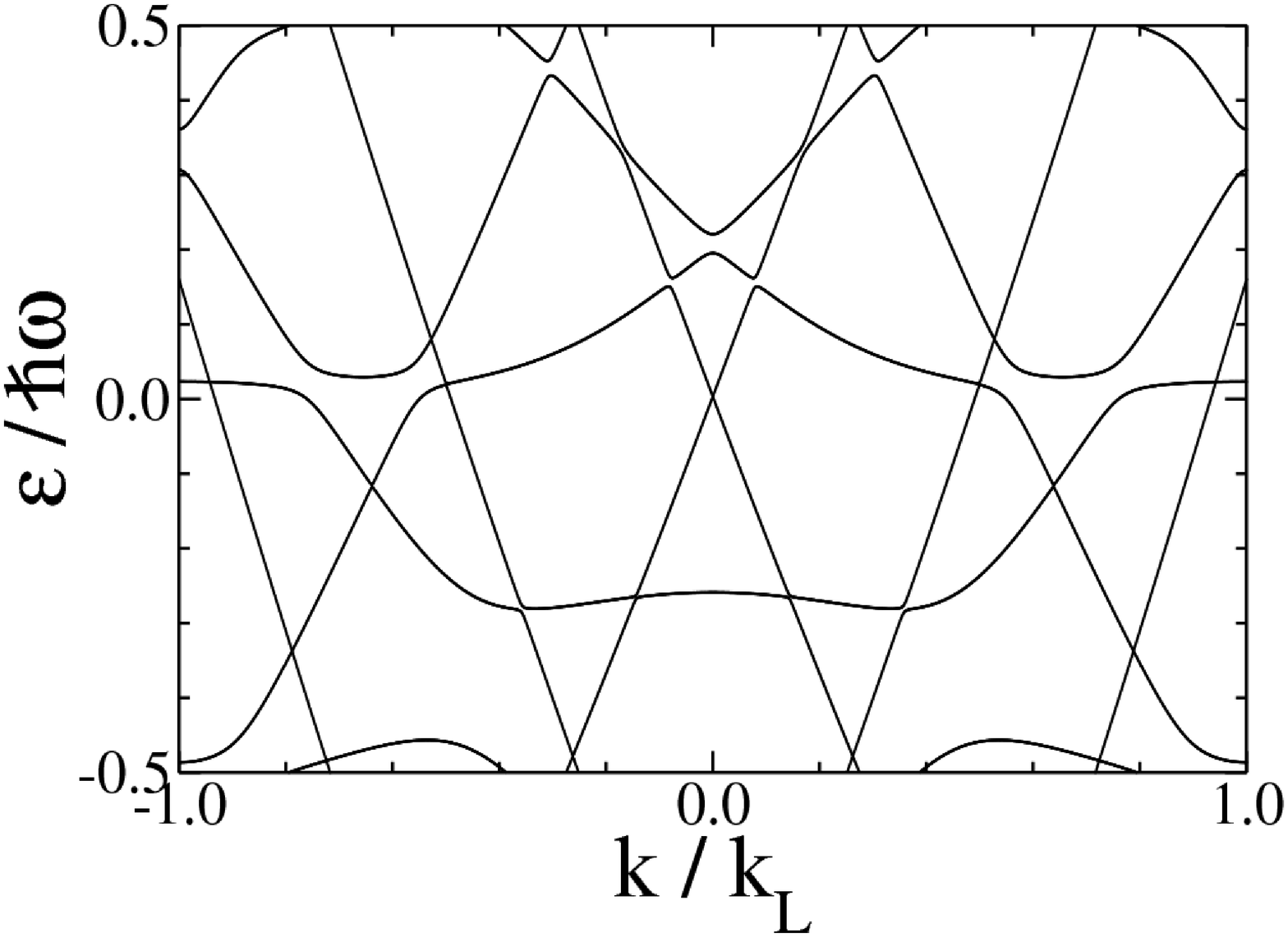}
\caption{Quasienergy dispersion relations for $V_0/\ER = 7.0$ and
	$\hbar\omega/\ER = 4.15$, providing a single-photon resonance
	between the lowest two energy bands. Scaled driving amplitudes
	are $\beta = 0.1$ (upper panel), $\beta =0.5$ (middle panel), and 
	$\beta = 1.0$ (lower panel). Observe how the wide double-well 
	dispersion created with $\beta = 0.5$ from the ground-state band is 
	punctured by narrow resonances when $\beta = 1.0$.} 	 
\label{F_14}	
\end{figure}

In the previous example multiphoton resonances have been considered to be
detrimental, causing unwanted transitions that should be avoided. But such
resonances also constitute one of the most powerful tools of Floquet
engineering: Resonances not only shift quasienergy curves, but they also 
break them up and create new connections. 
Seen from a conceptual viewpont, usual energy 
bands result from (energetically) ``horizontal'' hybridization, that is, by 
forming Bloch bands from energetically aligned Wannier states localized at 
the different lattices sites~\cite{Kohn59}. In contrast, quasienergy band 
formation can also involve ``vertical'' hybridization when different Wannier 
states localized at the {\em same\/} site are efficiently coupled by the 
oscillating force, as is achieved by choosing a matching frequency. This 
option, which has no counterpart in standard solid-state physics, can be 
exploited to design dispersion relations not found with any time-independent 
lattice. To illustrate this decisive feature, we again consider lattices with 
depth $V_0/\ER = 7.0$, as in Fig.~\ref{F_11}, but now with the lower driving 
frequency $\hbar\omega/\ER = 4.15$, which equals the speparation of the lowest 
two energy bands at $k/\kL \approx \pm 0.41$. As seen in the upper panel of 
Fig.~\ref{F_14}, this single-photon resonance causes a relatively wide 
anticrossing of the bands $n = 0$ and $n = 1$ even for a driving amplitude as 
small as $\beta = 0.1$, while the bands $n = 1$ and $n = 2$ undergo an 
anticrosing at $k/\kL \approx  \pm 0.55$. With increasing driving strength 
these avoided band crossings become wider, leading at $\beta = 0.5$  
(middle panel) to a double-well dispersion with minima spaced by  
$\Delta k/\kL \approx 0.9$, about twice as wide as those previously displayed 
in Fig.~\ref{F_12}. This ability to purposefully engineer such wide 
double well-shaped dispersion relations may be of some experimental interest: 
While the condensate wave function collapses into one of the two minima when 
these minima are close~\cite{ParkerEtAl13,HaEtAl15}, this might no longer be 
so when they are farther apart, possibly providing information on the 
condensate's coherence. But again, one has to take care of additional unwanted 
narrow resonances with driven free particle-like above-barrier states which 
inevitably will show up under strong driving: As shown in the lower panel of 
Fig.~\ref{F_14}, two such resonances puncture the double well right at its 
bottom when $\beta = 1.0$, opening up efficient channels of escape which might 
not be desirable.

\begin{figure}[t]
\includegraphics[width = 0.9\linewidth]{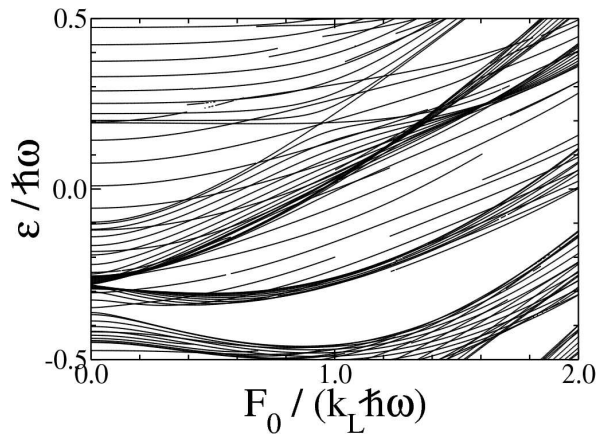}
\caption{Quasienergy bands for $V_0/\ER = 7.0$ and $\hbar\omega/\ER = 4.15$
	(cf. Fig.~\ref{F_14}). As in Fig.~\ref{F_13}, only quasienergy 
	representatives emanating from the lowest three energy bands are 
	plotted; each such band is represented by 21 equally spaced scaled 
	wave numbers $k/\kL$. Observe that there appear to be more than three 
	quasienergy bands in the strong-driving regime.}
\label{F_15}	
\end{figure}

But there is still more. If one inspects the development of the quasienergy
bands with increasing driving amplitude under these resonant conditions, as
displayed in Fig.~\ref{F_15}, one finds that the assignment of band indices
in the strong-driving regime is anything but obvious; as a result of the
resonant piercing of one band by another, additional bands seem to appear. 
Such a reorganisation of the band structure with increasing driving strength
necessarily gives rise to highly unusual dynamics of Floquet wave
packets~(\ref{eq:WPA}) populating an individual quasienergy band. The 
possibility to systematically engineer, experiment with, and utilize such 
exotic band structures may well constitute one of the greatest new chances 
offered by periodically driven optical lattices.

\section{Conclusions and outlook}
\label{S_5}

The preceding study has led to some computational insights which deserve to 
be formulated here, and to some more general conceptual issues which may 
inspire future experiments.

To begin with, the quasienergy eigenvalue problem for a driven optical lattice
is quite similar to that for the driven particle in a box discussed in 
Chap.~\ref{SubSec33}, except that the wave number~$k$ appears as an additional 
parameter for given values of the driving frequency and driving amplitude. 
Since the external force, when built into the optical-lattice Hamiltonian 
in one of the equi\-valent forms~(\ref{eq:HDC}), (\ref{eq:HXT}), 
or~(\ref{eq:HSH}), does not couple unperturbed Bloch waves with different wave 
numbers, and since the energy of Bloch states with the same wave number grows 
about quadratically with the band index, efficient coupling occurs only between
a limited number of states, greatly facilitating the numerical task for any 
fixed wave number. But still, one has to keep track of all wave numbers within 
one quasimomentum Brillouin zone to map out the band structure. Therefore, the 
resulting quasienergy bands may give rise to fairly complicated dynamics, as 
exemplified by Fig.~\ref{F_7}. Here the concept of coarse graining comes into 
play: If one may safely neglect all ``small'' quasienergy anticrossings, it 
may be possible to describe the driven system by an effective time-independent 
Hamiltonian within the subspace considered~\cite{GoldmanDalibard14}. The 
viability of this concept is illustrated by Fig.~\ref{F_9}: While it is still 
not possible to construct the full time-independent operator~$G$ which governs
the driven optical lattice in the sense of the transformed Schr\"odinger 
equation~(\ref{eq:SEG}), one may restrict oneself to the dynamics within 
the lowest undriven energy band under the conditions of Fig.~\ref{F_9}, and 
describe these dynamics in terms of an effective time-independent Hamiltonian
possessing only one single cosine energy band~(\ref{eq:IQB}), ignoring the 
$\bmod \; \hbar\omega$-indeterminacy of the original quasienergies. Whether 
or not such a coarse graining approach is feasible also depends on the time 
scale of the experiment. A small anti\-crossing of width $\delta\varepsilon$ 
may reasonably be ignored on time sales which are short compared to 
$\hbar/\delta\varepsilon$, but will make itself felt on longer time scales. 
Thus, it should be of interest to perform detailed measurements of the times 
required for heating, or simply for loss of particles from a driven optical 
lattice. Expressed differently, systematic measurement of heating rates
of ultracold atoms in driven optical lattices constitutes a means of
quasienergy band spectroscopy.

The experimental exploration of the close connection between avoided crossings 
of quasienergies and multiphoton resonances would be most illuminating in 
situations in which only relatively few, large anticrossings dominate the 
dynamics. Such situations can be deliberately engineered by vertical 
hybridization, in the sense explained when discussing Fig.~\ref{F_14}. The
future study of such multiphoton resonances with Bose-Einstein condensates 
in driven optical lattices would allow one to systematically address the 
influence of interparticle interactions on multiphoton transitions, with 
a clarity which probably will remain unachieveable in experiments with
crystalline solids exposed to laser radiation, thus creating an altogether 
new crosslink between the dynamics in strong laser fields and the physics of 
ultracold atoms. The recently reported, groundbreaking first experimental 
results obtained along these lines~\cite{WeinbergEtAl15} suggest that this 
optimistic view may not be unjustified. 

Such optical-lattice experiments aiming to utilize multiphoton-like transitions
with ultracold atoms will gain additional handles of control if they are not 
performed simply with long ac driving, but rather with short pulses possessing 
a smooth envelope. In such cases the quasienergy bands $\varepsilon_n(k)$, 
considered for all instantaneous driving strengths, form ``quasienergy 
surfaces'' on which the system's wave function can evolve in an effectively 
adiabatic manner as long as it does not encounter large avoided crossings, 
whereas it undergoes Landau-Zener-like transitions at such anticrossings, 
similar to the pulsed particle in a box examined in Fig.~\ref{F_4}. But with 
pulsed optical lattices such anticrossings would couple quasienergy surfaces 
only in the vicinity of certain wave numbers, as illustrated by Fig.~\ref{F_14}.
Therefore, one could exploit such anticrossings with driven optical lattices 
for selectively manipulating only certain components of given wave packets, 
leaving others unaffected~\cite{ArlinghausHolthaus11b}. Such pulse experiments 
could be given a further level of sophistication if they were combined with the 
feedback-techniques of optimal control, as they are used in chemical physics
in order to design specific laser pulses for the coherent control of molecular 
dynamics~\cite{JudsonRabitz92,RiceZhao00,ShapiroBrumer03}. In this manner, 
one could iteratively adapt the pulses' properties such that they result,
for instance, in the preparation of particularly interesting target states 
which may not be accessible by more conventional means: Instead of
``teaching lasers to control molecules''~\cite{JudsonRabitz92}, one then
teaches optical lattices to control Bose-Einstein condensates.
    
Ultracold atoms in optical lattices so far have created an immensely fruitful 
link between experimental atomic physics and the physics of condensed-matter 
systems~\cite{MorschOberthaler06,BlochEtAl08,LewensteinEtAl12}. It stands to
reason that the further investigation of ultracold atoms in time-dependent
optical lattices, driven either periodically or by pulses, will add a new
dimension to this field.

\begin{acknowledgments}
I would like to express my gratitude to all colleagues and collaborators
who have taught me many of the subjects collected here, with particular 
thanks to Heinz-Peter Breuer, Klaus Drese, Andr\'{e} Eckardt, and Stefan 
Arlinghaus. This work was supported by the Deutsche Forschungsgemeinschaft 
(DFG) through grant No.\ HO 1771/6-2. 
\end{acknowledgments}

\appendix

\section{Use of symmetries}

The most time-consuming step in the numerical computation of Floquet states 
and their quasienergies is the determination of the monodromy matrix. By
exploiting the symmetries of the quasienergy operator for a sinusoidally
driven optical cosine lattice, the numerical effort is reduced by a 
factor of two.

Let $\widetilde{H}(z,\tau)$ denote the scaled Hamiltonian appearing on
the right-hand side of Eq.~(\ref{eq:NUM}), multiplied by $\ER/(\hbar\omega)$,
so that the scaled quasienergy eigenvalue equation for the driven lattice 
takes the form
\begin{equation}
	\left( \widetilde{H}(z,\tau) - \ri \frac{\rd}{\rd \tau} \right)
	u(z,\tau) = \frac{\varepsilon}{\hbar\omega} u(z,\tau) \; ,
\end{equation}
with one period of the dimensionless time $\tau$ having the length $2\pi$.
The quasienergy operator here remains invariant under the combined operation
\begin{equation}   
	\widetilde{P} :	\left\{ \begin{array}{l}
		z 	\to 	-z		\\
		\tau	\to 	\pi - \tau	\\
		\mbox{complex conjugation}	
		    	\end{array} \right.	
\end{equation}
which implies 
\begin{equation}
	u(z,\tau) = u^*(-z,\pi - \tau)
\end{equation}
for each Floquet eigenfunction. In particular, this gives the identities 
\begin{eqnarray}
	u(z,\pi/2) 	& = & u^*(-z,\pi/2)
\nonumber \\
	u(z,0)		& = & u^*(-z,\pi) \; .
\label{eq:SYM}
\end{eqnarray}
Now the time-evolution operator for the first quarter cycle, ranging
from $0$ to $\pi/2$, is written as   

\begin{equation}
	\widetilde{U}(\pi/2,0) = \sum_n u_n(z,\pi/2) \, u_n^*(z',0) \, 
	\re^{-\ri \alpha_n} \; ,
\end{equation}	
where 
\begin{equation}
	\alpha_n = \frac{\varepsilon_n}{\hbar\omega} \frac{\pi}{2}
	= \frac{\varepsilon_n T/4}{\hbar}
\end{equation}
abbreviate the accompanying phases. Using the notation 	
\begin{equation}
	\varphi_\mu(z) = \langle z | \mu \rangle
\label{eq:TBF}
\end{equation}	
for the trigonometric basis functions~(\ref{eq:BA0}) and (\ref{eq:BAR}),
one thus has the matrix elements
\begin{equation}
	\widetilde{U}(\pi/2,0)_{\mu\nu} = \sum_n 
	\langle \mu | u_n(\pi/2) \rangle \langle u_n(0) | \nu \rangle  \, 
	\re^{-\ri \alpha_n} \; .
\end{equation}	
Next, the evolution operator for the second quarter cycle reads
\begin{eqnarray}
	\widetilde{U}(\pi,\pi/2) & = & \sum_n u_n(z,\pi) \, u_n^*(z',\pi/2) \, 
	\re^{-\ri \alpha_n}
\nonumber \\	& = & \sum_n u_n^*(-z,0) \, u_n(-z',\pi/2) \,
	\re^{-\ri \alpha_n} \; ,
\end{eqnarray}	
where the symmetries~(\ref{eq:SYM}) have been used. Observing now that
the basis functions~(\ref{eq:TBF}) have (ordinary) parity $(-1)^{\mu}$,
this gives   
\begin{eqnarray}
	\widetilde{U}(\pi,\pi/2)_{\mu\nu} & = & (-1)^{\mu + \nu} \sum_n
	\langle u_n(0) | \mu \rangle \langle \nu | u_n(\pi/2) \rangle \,
	\re^{-\ri \alpha_n}
\nonumber \\	& = &
	(-1)^{\mu + \nu } \, \widetilde{U}(\pi/2,0)_{\nu\mu} \; .	
\end{eqnarray}
Thus, having computed the evolution matrix for the first quarter cyle
one deduces its continuation for the second quarter cycle by symmetry; 
a corresponding identity connects $\widetilde{U}(2\pi,3\pi/2)$ and
 $\widetilde{U}(3\pi/2,\pi)$.

\end{document}